\documentclass[]{spie}  

\usepackage{amsmath,amsfonts,amssymb}
\usepackage{float}
\usepackage{graphicx}
\usepackage[colorlinks=true, allcolors=blue]{hyperref}
\usepackage{graphicx}
\usepackage{hyperref}
\usepackage[capitalise]{cleveref}
\usepackage{tabularx}
\usepackage{multicol}
\usepackage{stfloats}
\usepackage{multirow}
\usepackage{booktabs}
\usepackage{subcaption}
\usepackage{threeparttable}
\usepackage{placeins}
\usepackage{afterpage}
\usepackage{todonotes}
\usepackage{verbatim}

\title{The Atacama Large Aperture Submillimeter Telescope (AtLAST): enabling large-scale sub-mm science beyond 2030}

\author[a]{Claudia Cicone}
\author[b,c]{Tony Mroczkowski}
\author[d,e,f]{Evanthia Hatziminaoglou}
\author[g]{Matthias Reichert}
\author[h]{Sabrina Sartori}
\author[i]{Amelie Saintonge}
\author[j]{Pamela Klaassen}
\author[k]{Francisco Montenegro-Montes}
\author[g]{Martin Timpe}
\author[g]{Aleksej Kiselev}
\author[g]{Danilo Költzsch}
\author[f]{Carlos De Breuck}
\author[f]{Pamela Pizarro}
\author[f]{Eelco van Kampen}
\author[l]{Tinus Stander}
\author[i]{Dirk Muders}
\author[r,i]{Yiqing Song}
\author[m]{Kotaro Kohno}
\author[n]{Sergio Poppi}
\author[o]{Yoichi Tamura}
\author[g]{Stefan Thoms}
\author[p]{Hans Kaercher}
\author[a,q]{Sven Wedemeyer}
\author[n]{Alessandro Attoli}
\author[a]{Stephen Molyneux}
\author[h]{Bruno Silva}
\author[h]{Guillermo Valenzuela Venegas}
\author[h]{Paola Velasco-Herrejón}
\author[h]{Marianne Zeyringer}
\author[h]{Yogesh Kumar Yadav}
\author[j]{Mark Booth}
\author[f]{Carlos Alberto Dur\'an}
\author[r]{Lars-Ake Nyman}
\author[b]{Serni Ribó}
\author[b,s,c]{Francisca Kemper}
\author[g]{Manuel Groh}
\author[h]{Leandro Aggio}
\author[h]{Mathias Hudoba de Badyn}
\author[t]{Benjamin Magnelli}
\author[a,q]{Mats Kirkaune}
\author[u]{Akira Endo}

\affil[a]{Institute of Theoretical Astrophysics, University of Oslo, Blindern 0315 Oslo, Norway}
\affil[b]{Institute of Space Sciences (ICE-CSIC), Cerdanyola del Vallès, Barcelona, Spain}
\affil[c]{Institut d'Estudis Espacials de Catalunya (IEEC), E-08860 Castelldefels, Barcelona, Spain}
\affil[d]{Instituto de Astrofísica de Canarias (IAC), 38205 La Laguna, Tenerife, Spain}
\affil[e]{Departamento de Astrof\'{i}sica, Universidad de La Laguna, 38206 La Laguna, Tenerife, Spain}
\affil[f]{European Southern Observatory (ESO), Karl-Schwarzschild-Str.\ 2, Garching 85748, Germany}
\affil[g]{OHB Digital Connect, Weberstra\ss e 21, D-55130 Mainz, Germany}
\affil[h]{Department of Technology Systems, University of Oslo, Norway}
\affil[i]{MPIfR, Bonn, Germany}
\affil[j]{UK Astronomy Technology Centre, Royal Observatory Edinburgh, Edinburgh, UK}
\affil[k]{IPARCOS-UCM, Universidad Complutense de Madrid, Madrid, Spain}
\affil[l]{Department of EEC Engineering, University of Pretoria, South Africa}
\affil[m]{University of Tokyo, Japan}
\affil[n]{INAF Osservatorio Astronomico di Cagliari, Italy}
\affil[o]{Nagoya University, Japan}
\affil[p]{Independent consultant}
\affil[q]{Rosseland Centre for Solar Physics, University of Oslo, Blindern 0315 Oslo, Norway}
\affil[r]{ESO, Vitacura, Santiago, Chile}
\affil[s]{ICREA, Pg. Lluís Companys 23, E-08010 Barcelona, Spain}
\affil[t]{Université Paris-Saclay, Université Paris Cité, CEA, CNRS, AIM, Gif-sur-Yvette, France}
\affil[u]{TU Delft, The Netherlands}

\authorinfo{Send correspondence to: claudia.cicone@astro.uio.no}

\begin{document} 
\maketitle

\begin{abstract}
The Atacama Large Aperture Submillimeter Telescope (AtLAST) is designed to be the largest (sub-)millimeter single-dish astronomical observatory and the first climate-neutral modern research infrastructure. This dual ambition aligns with the objectives of the European Union, which funded its design study and the current design consolidation phase. AtLAST offers a unique combination of large aperture (50 m), large field of view (FoV; $>1$ deg), fast scanning speed (up to 3 deg/s), and high surface accuracy (20$\mu$m nighttime half wavefront error) that allows $\geq50$\% Ruze efficiency up to $\sim1$~THz. The design features a rocking chair mount with an active main reflector surface, a high precision closed-loop metrology system, and the space to house six major instruments. Instruments will be periodically updated as spectroscopic focal plane array, detector, coherent amplifier, and semiconductor technologies used in readout and backend electronics will significantly advance over the next decades. 
AtLAST will be a multi-purpose facility and is set to produce transformational results spanning nearly all fields of Astrophysics, such as Astrochemistry, Galactic and Extragalactic Astronomy, Cosmology, Planetary science, Stellar and Solar Physics, High energy astrophysics, and Time domain astronomy.
AtLAST's unrivalled throughput of $\rm A\cdot\Omega\sim6170~m^2~deg^2$ will enable wide-field ($>500$~deg$^2$) unbiased surveys. These will overcome extragalactic confusion noise and enable the detection of normal galaxy populations out to $z=7$. 
AtLAST will also reveal and characterise the missing baryons in the Universe, by mapping the elusive, low surface brightness gas within and around galaxies across cosmic time.
This gas can be traced in both its cold ($T\leq100$~K) and warm/hot ($T>10^5$~K) phases through molecular and atomic line emission, and the thermal and kinetic Sunyaev-Zeldovich effects, respectively. 
AtLAST will be the first green off-grid observatory, powered by a bespoke renewable energy system and reusing its braking energy thanks to a cutting-edge energy recovery system. By sharing surplus power and technological know-how with local communities, AtLAST will contribute to energy justice in Chile. AtLAST's new bold vision of a sustainable pursuit of breakthrough astronomy is an exceptional opportunity to shape the future of scientific research infrastructures.
\end{abstract}
 

\section{INTRODUCTION}\label{sec:intro}  
Astronomical observations at (sub-)millimeter ((sub-)mm) wavelengths are uniquely suited for studying regions of the Universe that are too cold, dusty, distant, or hot and tenuous to be probed at visible or other wavelengths. Current (sub-)mm telescopes have achieved major breakthroughs, beyond what is possible with optical and infrared observations. 
However, many emerging science goals now exceed the capabilities of existing telescopes.
Achieving the next transformational leap in discovery requires a new, large-aperture, multi-purpose single-dish telescope with a wide field of view (FoV) and a fast mapping speed across the full (sub-)mm band (30-950 GHz). These requirements cannot be met through upgrades of existing facilities, but demand a fundamentally new approach to telescope and instrument design.

In parallel, climate change and fuel price volatility are challenging many fields, including astronomy, to become more sustainable and to responsibly use energy resources. Because they require clear skies, dry atmospheric conditions, and a pristine radio environment, major astronomical observatories are often located in remote areas, sometimes near communities with limited access to reliable power. Telescope operations are power intensive, and, particularly in the (sub-)mm/radio bands, they may continue up to 24 hours per day 
requiring a continuous year-round power supply. Historically, sustainability has not been a primary driver in observatory design and operations, leading to continuous dependence on  fossil fuels. 
Active research in the field of renewable energy, tailored to the needs of astronomical facilities, is therefore a necessary first step towards a meaningful transition.

The dual ambition to pursue transformational science and make astronomy sustainable has motivated the Atacama Large Aperture Submillimeter Telescope (AtLAST\footnote{\href{https://www.atlast.uio.no}{https://www.atlast.uio.no}}).
The EU-funded AtLAST design study (2021-2024\footnote{\href{https://cordis.europa.eu/project/id/951815}{https://cordis.europa.eu/project/id/951815}}, hereafter referred to as AtLAST1) advanced the facility from the conceptual stage to mid-point of the design phase \cite{Mroczkowski+25}. The design study was driven by the core values of scientific excellence, technological innovation, open science, collaboration, inclusivity, and environmental sustainability, with the goal of creating a next-generation facility for the astronomical community \cite{Klaassen+20}. 
AtLAST1 delivered the design for an environmentally sustainable single dish (sub-)mm telescope with a 50-m primary mirror and an unprecedentedly large (up to 2~deg diameter) FoV for this aperture size,\cite{Gallardo+24} hosting a suite of six instruments 
\cite{AtLAST_memo_3, Mroczkowski+23, Mroczkowski+25}. While 
some of these capabilities  exist individually on current or planned astronomical facilities, their combination in a single facility is unique to AtLAST and represents a significant technological and scientific advance \cite{Klaassen+20}. 
This achievement required original multi-disciplinary research, spanning different academic fields within the domains of Natural, Formal, and Applied Sciences  \cite{Mroczkowski+25, Reichert+24, Gallardo+24, Puddu+24, Viole+23, Viole+24a, Viole+24b, valenzuela2024renewable, Schimek+24a, Schimek+24b, Kirkaune+25}. 

The ongoing EU-funded AtLAST2 project, {\it Consolidating plans for the Atacama Large Aperture Submillimeter Telescope}, runs from 2025 through 2028\footnote{\href{https://cordis.europa.eu/project/id/101188037}{https://cordis.europa.eu/project/id/101188037}} and  involves 16 participant institutes from Europe and South Africa, and four associated partners from Japan. In addition to over 120 team members from the  AtLAST2 partners, about 50 scientists from other institutes across the globe joined the consortium through Non Disclosure Agreements. Building on this broad international expertise,
AtLAST2 will consolidate the telescope design by (i) demonstrating innovative technological features; (ii) reviewing the telescope system and its interface with the energy system; and (iii) influencing the development of diverse and multi-purpose first-light instrumentation concepts. 
Along with modern operations concepts that maximize efficiency and scientific throughput, AtLAST2 will also deliver a life cycle assessment of the facility including global aspects to further reduce its environmental impact, using existing facilities as test bed infrastructures. By the end of 2028, AtLAST will have increased the technology readiness level (TRL) of its crucial components and undergone preliminary design review, ready to move the project to its implementation phase. 

This paper provides an overview of the current status of AtLAST development, highlighting the latest design consolidation results and outlining the expected  progress through the remainder of AtLAST2.

\section{AtLAST: AN OVERVIEW}

 AtLAST will be a single-dish (sub-)mm observatory with a 50~m primary mirror, delivering a maximum angular resolution of $1.22\lambda/D\sim1.6''$ at $950$~GHz. It will host four 1-deg FoV instruments corotating in elevation, and two 2-deg FoV instruments mounted on Nasmyth platforms. All six instruments must be readily available on-demand for observations. AtLAST's maximum throughput of $\rm A\cdot\Omega\sim6160~m^2~deg^2$ is higher than any other (sub-)mm/mm telescope including Simons Observatory and the Fred Young Submillimeter Telescope, and it is almost 20 times larger than that of Vera C. Rubin Observatory.
The telescope is designed to sustain maximum scanning speeds of 3 deg/s and accelerations of 1 deg/s$^2$. 
These technical constraints are driven by transformational science goals (Sec~\ref{sec:science}). Satisfying them requires a new, cutting-edge design, inspired by 30-m class optical/infrared telescopes (Sec~\ref{sec:design}).
The most ambitious science goals demand AtLAST to observe across the full (sub-)mm range up to the highest frequencies accessible from ground 
($\sim950$~GHz, i.e.\ 316~$\mu$m, see Sec~\ref{sec:science}). For this, AtLAST needs a high ($\sim5000$~m) and dry site with excellent atmospheric transparency (Sec~\ref{sec:site}), and a surface accuracy of $\leq$~20$\mu$m nighttime half wavefront error, which will be relaxed to $\leq$~30$\mu$m at daytime when higher wind and precipitable water vapour (PWV) limit observations at the highest frequencies. Such high surface accuracy requires an active primary surface and a closed-loop live metrology system (Sec~\ref{sec:soap}), as well as a site shielded from the strongest winds (Sec~\ref{subsec:wind}).

AtLAST is conceived as a facility that will serve a wide scientific community for decades to come. It is therefore indispensable that multiple instruments can be installed and at the ready for observations, and that switching between them can be done quickly and safely, with an operational flexibility that can maximally exploit the excellent atmospheric conditions routinely available from the Chajnantor plateau. A ground-based multi-purpose astronomical observatory such as AtLAST enables continuous instrumentation upgrades and is the catalyst of significant technological advances. Its large FoV coupled with the small beam ensure that AtLAST will remain relevant and upgradable for many decades, since for many instrumentation technologies it will take some time to develop focal plane arrays with millions of elements on sky (Sec~\ref{sec:inst}). 

A bespoke renewable energy system has been designed for AtLAST to be environmentally sustainable (Sec~\ref{sec:renewable}). Ongoing research is focusing on developing and testing metal hydrides intermetallic alloys for storage tanks and compressors that are tailored for the specific
use of AtLAST (Sec~\ref{sec:MHs}). In parallel, the team is eliciting the views on sustainability and energy from different stakeholders in Chile, to make AtLAST an infrastructure that supports climate mitigation while contributing to the wellbeing of neighbouring communities (Sec~\ref{sec:socialjust}).

To maximise the science output, AtLAST places its users at the core of its operations plan (Sec~\ref{sec:operations}). An adequate data infrastructure is needed to: (i) store the large volumes of raw data expected for new state-of-the art instrumentation; (ii) enable efficient remote and distributed operations.

\section{SCIENTIFIC MOTIVATION}\label{sec:science}
\subsection{A leap in (sub-)mm survey and mapping capabilities}

\begin{figure}[tbp]
	\centering
	\includegraphics[width=\textwidth]{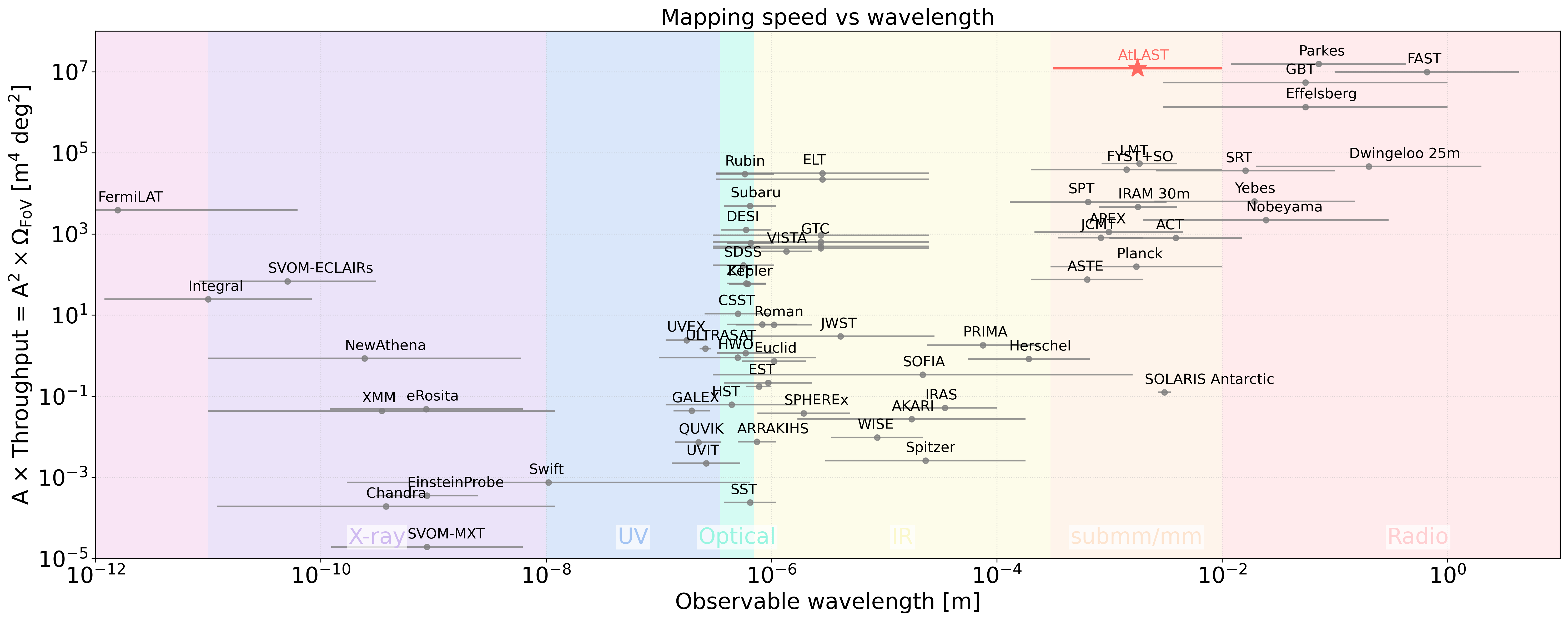}
	\caption{Collecting area multiplied by the throughput (or étendue), used as a proxy for the mapping speed, plotted as a function of wavelength for a selection of (single-antenna) astronomical facilities across the electromagnetic spectrum. This mapping speed metric, which scales with ${\rm A^2}$, is similar to the telescope figure of merit used elsewhere,\cite{Wilson2013, Mroczkowski+25} but does not include the bandwidths, which are generally limited by instrumentation before they are limited by the telescope design.  
    The following facilities are plotted but not labelled to minimise text overlap: Keck, LBT, VLT (single UT), GMT, DES, TESS, Gaia, Lazuli. The full observable wavelength range of each facility is displayed using the horizontal errorbar. The FoV size used in the computation is the maximum FoV enabled by the telescope design, which does not necessarily apply to the full wavelength range covered by the facility. For the X-ray and Gamma-ray telescopes, the effective area was used. 
    For the radio telescopes, we used the information on the size of the FoV enabled by multi-beam receivers, when available.
    Some of the facilities plotted are not yet approved.}
	\label{fig:mapping_speed}
\end{figure}

Our theoretical understanding of the Universe is grounded in our ability to survey large areas of the sky across multiple wavelengths. Wide-field ``blind'' surveys, spanning the full range of astrophysical environments and cosmic epochs, deliver statistical insights into whole populations of sources: protoclusters and clusters of galaxies, galaxies, stars, black holes, planets, comets, asteroids, etc. These large datasets are the primary testbeds of astrophysical theoretical models. The (sub-)mm wavelength range is not just a mere complement to the other wavebands: it has a unique value at probing matter that is invisible at other wavelengths, either because extremely cold (temperature, $T\leq10^2$~K), extremely warm/hot ($T\geq10^5$~K), extremely dusty and obscured, or because its bright line diagnostics are redshifted out of the ultraviolet, optical, and infrared bands. 

Some of the current and planned (sub-)mm facilities are able to map large areas of the sky thanks to their wide deg-scale FoVs, but the datasets they produce are shallow because of their small 6-10m apertures and resulting high confusion noise. None of these (sub-)mm facilities can match the survey depth and statistics that e.g.\ Rubin and the SKA will deliver at other wavelengths. 
AtLAST fills this gap: it will be the first widefield ($>1$~deg FoV) single dish observatory that is not confusion-limited at (sub-)mm frequencies, thanks to its high angular resolution ($\rm 1.22\lambda/D=1509^{\prime\prime}/\nu[GHz]$) and high sensitivity (1~mJy RMS in $<8$~s at 350~GHz in $\Delta\nu=8$~GHz channels). The latter is comparable to that offered by ALMA, but on AtLAST it would be accompanied by a mapping speed $10^{3}$--$10^{5}\times$ higher than that of ALMA, and by the ability to capture such faint signals even when arising from large angular scales up to $\sim 1~\rm deg^2$, which is not possible with (sub-)mm interferometers \cite{Booth+24_SPIE}. Figure~\ref{fig:mapping_speed} shows the leap in mapping speed enabled by AtLAST in the (sub-)mm. This plot allows a meaningful comparison of mapping speed between facilities at similar wavelengths, while providing an overview of the landscape of single-antenna telescopes operating across the electromagnetic spectrum. Cross-wavelength comparison of the ${\rm A^2\times\Omega_{FoV}}$ metric shown in Fig.~\ref{fig:mapping_speed} should be treated with caution, as this metric does not account for differing atmospheric noise, detector technologies, and science drivers. 

As presented in a series of peer-reviewed publications \cite{Booth+24_SPIE, 2024ORE.....4...78C, 2024ORE.....4..112K, 2024ORE.....4..117L, 2024ORE.....4..122V, 2024ORE.....4..140W, 2025ORE.....4..113D, 2025ORE.....4..148L, 2025ORE.....4..132O, Schimek+24a, Schimek+24b},
AtLAST is designed to produce transformational results in most fields of Astrophysics, such as Astrochemistry, Galactic and Extragalactic Astronomy, Cosmology, Planetary science, Stellar and Solar Physics, High energy astrophysics, and Time domain \& transient astronomy. For example, AtLAST would enable a $\sim$1,000 deg$^2$ (sub-)mm continuum survey detecting $\sim$54 million galaxies, resolving more than 80\% of the cosmic infrared background at $\lambda\leq750$~$\mu$m, sampling numerous Coma-like clusters out to $z\sim2$ and group environments out to $z\sim6$ \cite{2024ORE.....4..122V}.
Such extraordinary performance will stem not only from AtLAST’s unparalleled mapping speed, but also from its access to high frequencies $\nu\geq400$~GHz. Diffraction-limited resolutions as fine as $1.6''$ at 950~GHz will dramatically reduce confusion noise while providing robust positional priors for source identification at longer, more confusion-limited wavelengths. Access to wavelengths as short as $\sim300~\mu$m is also crucial for probing the peak of the dust emission in high-z galaxies and, when combined with mm coverage, for fully constraining their dust spectral energy distributions, thereby breaking the well-known degeneracy between dust temperature, emissivity, and infrared luminosity. 
By delivering dust-obscured star-formation rates and dust properties for normal star-forming galaxies out to $z\sim6$, this survey would provide the most precise measurement of the dust-obscured cosmic star-formation history over the past 13~Gyr and establish how environment shapes galaxy growth, quenching, and morphological transformation across cosmic time. Complementing this effort, a $\sim$1 deg$^2$ blind spectroscopic survey would characterize $\sim115,000$ normal star-forming galaxies out to $z\sim7$, providing secure redshifts together with key interstellar medium (ISM) diagnostics from multiple molecular and atomic lines \cite{2024ORE.....4..122V}. Together, these surveys can deliver the first statistically robust census of dust, gas, and star formation from the local Universe to the epoch of reionization, revealing both the physical drivers of galaxy evolution and the emergence of dust in the early Universe.

Another ambitious anticipated science goal is the discovery and study of the missing baryons in the Universe, i.e.\ the gaseous matter permeating galaxy haloes and the cosmic web that has so far escaped direct measurements. Most of the missing baryons are expected to be in a tenuous, warm-hot ($T>10^5$~K) gas phase, but a large fraction may also be hidden in a cold ($T<10^2$~K) neutral phase accessible only through atomic and molecular lines at (sub-)mm wavelengths. As demonstrated in Sec~\ref{sec:performance} (see also \cite{Bonanomi+24}), interferometers miss significant flux already on scales of 10s of arcsec, and such flux loss is particularly dramatic at high frequencies where it reaches values of $>90\%$. AtLAST, thanks to its unprecedented throughput, broad frequency coverage, and sensitivity to structures from arcsecond to deg-scales, can capture missing baryons in both the warm/hot and cold phases, using respectively the thermal (and kinetic) Sunyaev-Zeldovich effect\cite{2025ORE.....4..113D} and molecular/atomic line emission \cite{2024ORE.....4..117L}.

AtLAST will fulfil the (sub-)mm scientific needs of communities that have so far had to build their own dedicated facilities. Thanks to its solar capabilities, AtLAST will be able to perform fast, multi-frequency, full-polarisation full-disk scans of the Sun, enabling time-dependent 3D reconstructions of the thermal and magnetic structure of the solar chromosphere. Such observations will reveal the physical processes behind the dynamics and heating of the solar atmosphere, as well as the solar origins of space weather, with important implications for operational space-weather forecasting \cite{2024ORE.....4..140W, Wedemeyer+25_ESOWP}.
Because no other facility combines arcsecond-scale resolution, sub-mJy sensitivity, and a degree-scale FoV, AtLAST is poised to transform the field of (sub-)mm time-domain and transient astrophysics via its high-cadence (on timescales of seconds) and rapid response observations  \cite{2025ORE.....4..132O, Koljonen+25_ESOWP}.

\subsection{Performance characterisation of AtLAST, ALMA, and ACA}\label{sec:performance}

AtLAST can recover missing flux across the full (sub-)mm range, not only at the very large degree scales, but also at relatively small scales of 10s of arcseconds to few arcminutes. The latter are of high scientific impact for a large number of science cases, including the study of the interstellar, circumgalactic and intergalactic media\cite{2024ORE.....4..117L, 2025ORE.....4..148L, 2025ORE.....4..113D}, and therefore the characterisation of the missing baryons. Although such scales are formally feasible to image at the lowest (sub-)mm frequencies with current interferometers, a large fraction of the flux is nonetheless missed, hence biasing the scientific output of these observations in ways that are currently unknown and unaccounted for. At $\nu>400$~GHz, the flux loss in interferometric observations exceeds $90\%$ in the example shown below.

\subsubsection{Test case at $\sim230$~GHz}

\begin{figure}[tbp]
	\centering
	\includegraphics[width=\textwidth]{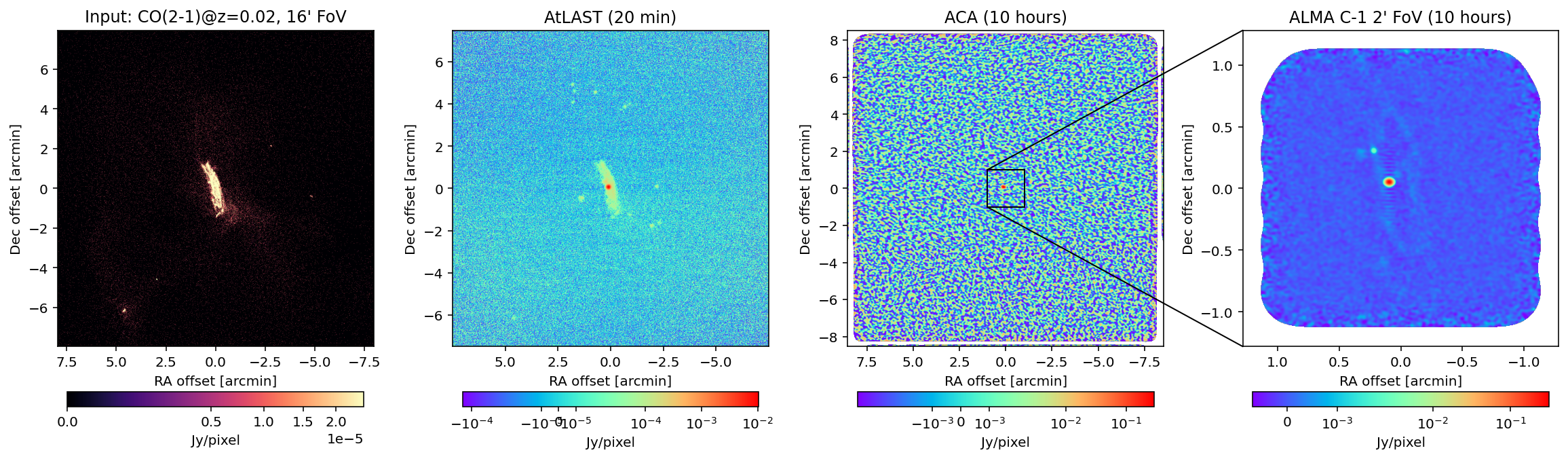}
	\caption{Performance of AtLAST, ACA, and ALMA at imaging the CO(2-1) molecular line emission from a massive star forming galaxy at $z=0.02$ across a $16'$~FoV (400~kpc). All mocks were produced assuming PWV=0.8~mm, $\rm \nu_{obs}^{CO(2-1)}=226.018$~GHz, and $\Delta\nu=0.753$~GHz (1000~km/s at $\rm \nu_{obs}^{CO(2-1)}$). The AtLAST mock image, obtained with the \texttt{maria} simulator, shows the result of a 20~min raster scan with a $14'$ FoV heterodyne array. The ACA and ALMA mock images, obtained with the \texttt{simobserve} and \texttt{simanalyse} tasks in \texttt{CASA}, show the result of mosaic observations with a total integration time of 10 hours each. In the ALMA simulation, the most compact array configuration was used (C-1), and the FoV was restricted to the central $2'\times2'$.}
	\label{fig:CO21_localgalaxy}
\end{figure} 

AtLAST's capabilities at $\sim230$~GHz are demonstrated in Fig.~\ref{fig:CO21_localgalaxy} through mock CO(2-1) maps of a massive, nearby ($z=0.02$) star-forming galaxy and its surrounding environment. The input image, drawn from the SIMBA cosmological simulation (credit: D. Narayanan \cite{Narayanan+20_VLAmemo}), represents the total velocity-integrated CO(2-1) map of a 400~kpc-size FoV ($16'$). The majority of the CO(2-1) luminosity ($\log L'_{\rm CO(2-1)} [\rm K~km/s~pc^2]=9.62$) arises from the central galaxy disk, which shows a bright nuclear region as well as more diffuse extended components. A few fainter dwarf companions are also noticeable in the input map. 
The AtLAST mock image shown in the second panel was obtained with the \texttt{maria} simulator\footnote{\href{https://thomaswmorris.com/maria/index.html}{https://thomaswmorris.com/maria/index.html}, \href{https://github.com/thomaswmorris/maria}{https://github.com/thomaswmorris/maria}}, with an $13.93'$ FoV instrument that mimics an heterodyne array with 10887 elements (e.g.\ feed horns, antennas, bare pixels) at 1 $f\lambda$ (1 beam) separation, assuming $T_{rx}=40$~K
and a total optical efficiency of $\eta=0.75$. The \texttt{maria} code adds atmospheric and detector noise and accounts for transmission losses \cite{Morris+22, vanMarrewijk+24}. To simulate the AtLAST observation, we used a raster scan with a $7'$ throw, 21 raster rows, a scanning speed of 5$'$/s, and a total observation duration of 20 min. The ACA and ALMA mock observations were obtained with the \texttt{simobserve} and \texttt{simanalyse} tasks in \texttt{CASA v6.7}, assuming an integration time of 10~hours and using the Cycle 13 configurations files. 85 pointings were required to map the full FoV with the ACA. For the ALMA mock, we used the most compact configuration (C-1) and restricted the FoV to the inner $2'$, which resulted in a mosaic map with 2125 pointings. The angular resolutions of the AtLAST, ACA, and ALMA C-1 maps are $6.7''$, $6.7''$, and $1.9''$, respectively. Figure~\ref{fig:CO21_localgalaxy} clearly shows that the target morphology, its molecular disk extent, and the presence of extended tails and dwarf companions, cannot be recovered after 10~h of ALMA or ACA observations, while AtLAST captures them in 20~min. Only the AtLAST map allows us to
observe the impact of gravitational interactions, feedback processes, and of galaxy group and cluster environments on the ISM of the target galaxy. 
Moreover, 
the ACA and ALMA mocks miss $\sim30\%$ of the input flux within a central aperture with $1'$ radius. The latter casts some doubts on our current ability to measure total molecular gas masses, even in very nearby massive galaxies. 

\subsubsection{Test case at 480~GHz}

\begin{figure}[tbp]
	\centering
	\includegraphics[width=\textwidth]{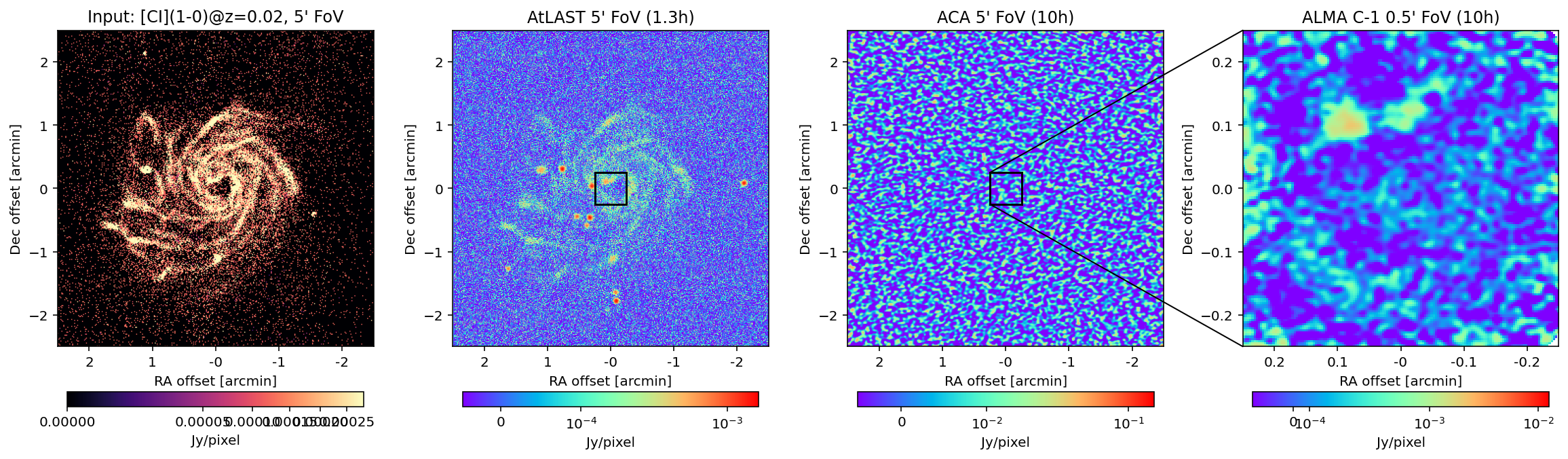}
	\caption{Similar to Fig.~\ref{fig:CO21_localgalaxy}, but in this case the input is the integrated [CI](1-0) line emission map, at $\rm \nu_{obs}^{[CI](1-0)}=482.510$~GHz, with a FoV of 5$'$ (120~kpc). All mocks use PWV=0.4~mm. The [CI](1-0) luminosity in the central $2'$ of the input image is $\log L'_{\rm [CI](1-0)} [\rm K~km/s~pc^2]=9.33$. The bandwidth is 1.609~GHz (1000~km/s at $\nu_{obs}^{[CI](1-0)}$). 
    For the AtLAST mock, we simulated with \texttt{maria} a 1.3h raster scan with a $5'$~FoV heterodyne array. The ACA and ALMA mock images show the result of 10h mosaic observations. For the ALMA C-1 simulation, the FoV was restricted to the central $30''\times30''$, shown as a black square in the second and third panels.}
	\label{fig:CI10_localgalaxy}
\end{figure} 

The flux loss is exacerbated at higher frequencies, as shown by Fig.~\ref{fig:CI10_localgalaxy}. Here, the input is the same simulated galaxy as in the low frequency case, but shown with a different orientation and with a smaller $5'$ FoV. The target line is the lowest energy atomic carbon transition, [CI]$^3P_1-^3P_0$ (hereafter [CI](1-0)), a tracer of cold molecular gas alternative to CO. This line holds the potential of revealing CO-dark H$_2$ gas reservoirs exposed to UV radiation and Cosmic rays which may be depleted of CO, but it is relatively unexplored at low redshift due to the low atmospheric transmission at $\nu\sim450-490$~GHz. 
For the AtLAST mock, we defined in \texttt{maria} a $5'$~FoV array instrument with 1609 spatial elements at 2 $f\lambda$ separation (typical of feed horn coupled detectors), assuming $T_{rx}=100$~K and optical efficiency $\eta=0.4$. We used a raster scan with $5'$ throw, 21 raster rows, scanning speed of $10'$/s, and a total observing time of 4800~s (1.3h). The ACA and ALMA mocks required 756 and 20 mosaic pointings to image respectively the $5'$ FoV and a $0.5'$ FoV in 10~h. 
The angular resolutions of the AtLAST, ACA, and ALMA C-1 maps are $3.3''$, $3.0''$, and $0.9''$. Figure~\ref{fig:CI10_localgalaxy} shows that only the AtLAST mock captures the atomic carbon emission from the input galaxy, including all its disk substructures. The ACA result is consistent with a non detection, and the ALMA mock displays only two bright compact regions near the nucleus of the galaxy, missing $>90$~\% of the input flux. Hence, without AtLAST, our ability to probe CO-dark H$_2$ gas through the [CI](1-0) line in the nearby Universe is severely jeopardised.

\section{TELESCOPE DESIGN}\label{sec:design}

\begin{figure}[tpb]
    \centering
    \vspace{-8mm}
    \includegraphics[width=0.85\linewidth]{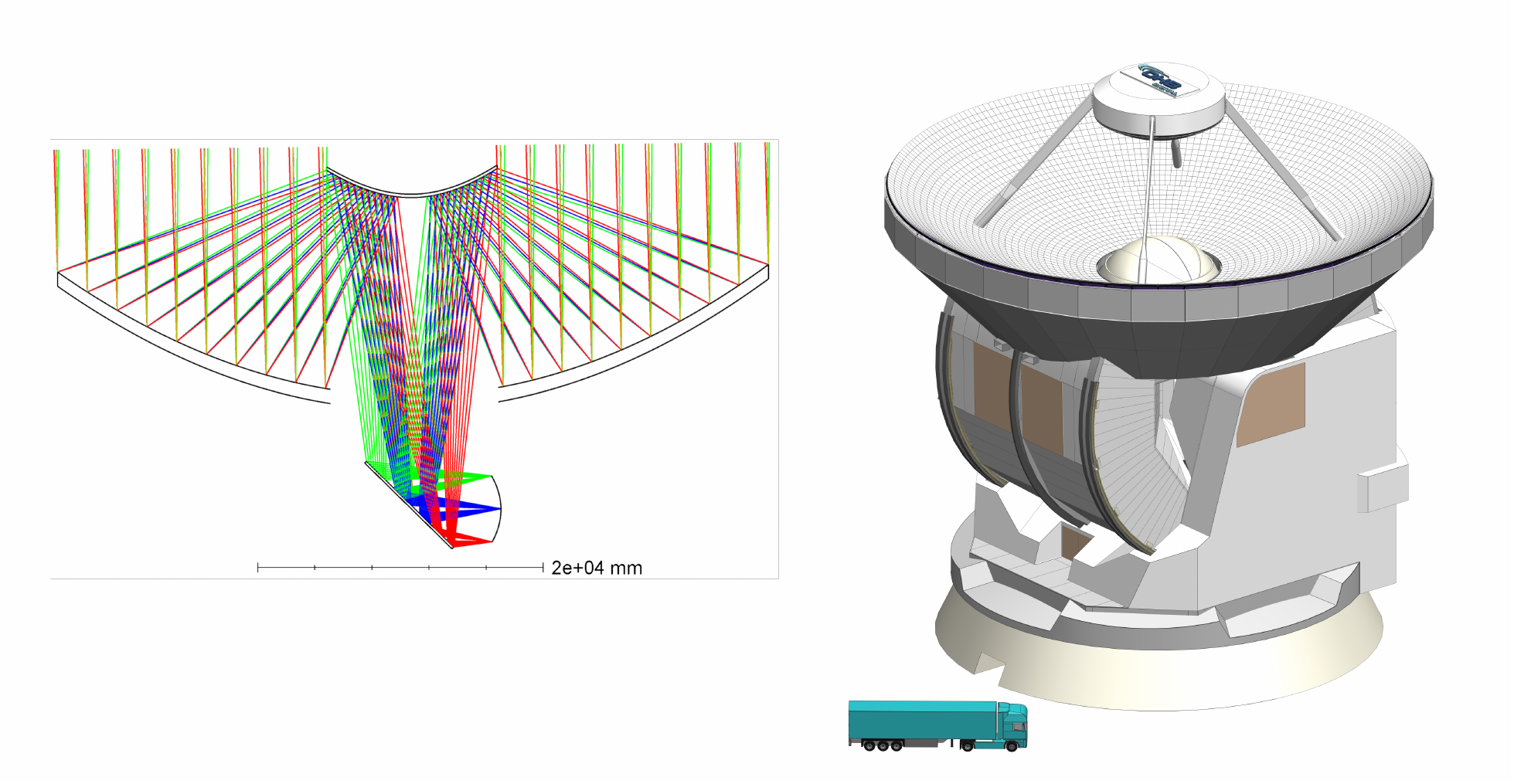}
    \caption{AtLAST optical layout ray trace (left) as presented in Gallardo et al.\ 2024\cite{Gallardo+24} and the structural design (right) concept as in Reichert et al.\ 2024\cite{Reichert+24}.  The fast optics and large secondary mirror shown on the left allow for the large focal surface. The rocking chair mount, shown on the right, enables a relatively compact design and large enclosure for instrumentation.}
    \label{fig:OpticsCAD}
\end{figure}

The conceptual design of AtLAST was presented in previous publications\cite{Reichert+24, Mroczkowski+25}; here we  briefly summarise the main features. AtLAST's optical design is based on a slightly modified Ritchey–Chrétien configuration, as shown on the left in Fig.~\ref{fig:OpticsCAD}. The hyperbolic primary mirror (M1) has a focal ratio of approximately $0.35$. The hyperbolic secondary mirror has a diameter of 12~m \cite{Gallardo+24}.
The beam is redirected by the tertiary folding mirror (M3) to one of two instruments located at the Nasmyth foci or, alternatively, to one of four instruments co-rotating in elevation at the Cassegrain foci. The nominal total focal length of the system is 134.608~m, corresponding to an overall focal ratio of 2.69.
The telescope is supported by a structure that includes an elevation system designed in the form of a rocking chair. This configuration provides sufficient space around the elevation axis to accommodate the Nasmyth instruments, which have an envelope of approximately 5~m in diameter. The rocking-chair structure moves on curved tracks supported by wheels, following a track-on-wheel design. This represents a reversed configuration of the conventional wheel-on-track concept typically used for the azimuth axis. The azimuth structure itself follows a conventional wheel-on-track design.

\subsection{Conceptual design review}\label{sec:CoDR}

AtLAST2, the design consolidation phase,
started with an independent review of the conceptual design to determine which aspects required additional effort beyond the expected evolution. 
To that end, a team of engineers, telescope designers, observers, and instrument experts who were not part of AtLAST1  was assembled in mid-2025. The review was successful, with the consensus being that the conceptual design was very advanced and constituted a strong basis for the consolidation phase.  Most recommendations were well-aligned with the development already prioritized by the AtLAST2 proposal e.g.\ optical alignment, metrology, and control systems, described in the next sections. 
In addition, the review panel recommended putting in place a formal framework for tracking project documentation, including engineering and technical documentation from the project as a whole, in a way that these documents could remain findable and accessible long after the end of the EU project. With this recommendation in mind, efforts are being undertaken to strengthen system engineering capabilities, with a focus on integrating mechanical, optical, and operational requirements into a unified design approach.

\subsection{System for Optical Alignment, Active Surface and Pointing}\label{sec:soap}
Given the large physical dimensions of the telescope structure -- a reflector diameter of 50~m, an azimuth track diameter of $\sim35$~m, a height of the elevation axis of $\sim25$~m and with at total mass estimate of $\sim4500$~t -- AtLAST must fulfil tight error budgets with respect to half wavefront error and pointing \cite{Reichert+24}. 
The large structural dimensions and mass result in considerable deflection amplitudes. In steel structures of this size and mass, variations in the gravity vector can produce deflections of up to several cm. If these effects are not compensated, they would lead to half-wave front errors and pointing errors that are several orders of magnitude above the specified requirement of $\leq\,20\mu m$ nighttime half wavefront error, $\leq 2.5''$ blind pointing accuracy and $\leq 0.5''$ rms over 10 minutes tracking accuracy at sidereal tracking speed. 
The main loads acting on the structure include external loads, such as gravity, wind, and varying temperature distributions, as well as internal loads caused by accelerations during scanning. Only a small fraction of the external load vectors is repeatable. 
Finite element analyses have shown that with passive measures a primary reflector surface accuracy of only $\sim200\,\mu m$ can be achieved not taking into account errors introduced by misalignment and other surfaces. And the blind pointing accuracy would be limited to approximately several dozen arcseconds. 
For these reasons, the primary reflector is developed at the current project state with an active surface to compensate local distortions of the reflecting surfaces shapes. In optical telescopes, the corresponding compensation method is known as active optics.
The major reflecting surfaces and the instruments will be continuously maintained in alignment with the optical axis. We foresee active control of deflections in the lateral, tilt, and piston directions. An active system will measure, as close to real time as possible, the translations and rotations of the reflecting surfaces and instruments. These motions are treated as rigid-body motions by using a best-fit representation of the M1 surface, M2, M3, and the instrument compartment. Such active alignment systems are well established technology and essential for  the optical/NIR 8-10~m class telescopes currently  in operation (e.g.\ VLT, Gemini, GTC, Keck) and the next generation 30~m class telescopes (ELT, TMT, GMT).

For both the alignment-error compensation and the correction of local surface deformations, the sensor and compensation technologies established in optical telescopes — such as wavefront sensors, active surfaces, and mirror cells — are not available for radio telescopes at the TRL required for AtLAST.
AtLAST will be directly exposed to environmental conditions, as it will not be protected by a dome. 
The majority of the scientific observations will require scanning motions of the telescope of some form. 
For the fastest scans, which we expect to be a smaller subset of the total scanning operations, the structure may experience angular velocities up to 3~deg~s$^{-1}$ and accelerations up to 1~deg~s$^{-2}$, which are very high compared to normal tracking trajectories of celestial objects or even routine mapping observations.
The optical alignment, active surface, and pointing control systems are therefore a key enabling technologies for AtLAST. 
No comparable system has been realized in a radio/(sub-)mm telescope to date, but it is indispensable for AtLAST.
A separate work in these SPIE proceedings by Kiselev et al.\cite{Kiselev+26_PE}  models the impacts of both high wind loading (using the real data described in Sect~\ref{sec:site}) and fast scanning on the point accuracy for AtLAST. The main findings are: (i) the residual pointing errors associated with high winds (15~m~s$^{-1}$) are acceptable and (ii) the pointing errors during fast scanning are deterministic and can be characterised in the telescope encoder values, meaning that the pointing can be recovered using standard mapmaking procedures \cite{Dunner2013}.

Two major aspects of the system for optical alignment, active surface control, and pointing are being studied in AtLAST2. Firstly, we are   
investigating suitable sensor technologies through:
(i) wavefront sensing, using the 45-m Nobeyama telescope as a technology demonstrator at mm-wavelengths \cite{Tamura2020}; and
(ii) the implementation of highly accurate distance measurement systems in an exposed environment, using the 64-meter Sardinia Radio Telescope (SRT) as a testbed, where a metrology system based on Hexagon’s Etalon sensors is currently being implemented \cite{Attoli2023}. The second aspect defines the breakdown of the wavefront and pointing error budgets. It comprises an assessment, based on analysis and simulation, of how control loops can convert sensor information into appropriate actuator motions in order to achieve the required speed and accuracy of compensation.
This work defines:
(i) tolerance ranges for alignment errors of the reflecting surfaces for each degree of freedom,
(ii) the minimum number of actuators required for an active M1 surface to compensate local deformations,
(iii) the minimum stroke and positioning accuracy required for the M1 surface actuators,
(iv) suitable actuator arrangements and a mechanical interface between the backup structure (BUS) and the panel segments \cite{Thoms+26_WFDSensorPlacement} and 
(v) a compensation control loop that enables rapid identification of deformation states and correction of alignment errors and local flexure of the reflecting surfaces under transient disturbances.
In the case of optical alignment, the alignment state is also coupled to the pointing direction. Alignment errors affect the wavefront error and the pointing accuracy. Therefore, both quantities are treated as coupled objectives. These design activities are a work in progress at OHB DC. 

\subsection{Energy recovery system}\label{sec:ERS}

A major contributor to the dynamic electrical load of AtLAST is the motion of the telescope structure. 
The combination of a 50-m aperture, a massive steerable mount, and the requirement for fast scanning motions with angular velocities up to 3~deg~s$^{-1}$ and accelerations up to 1~deg~s$^{-2}$ leads to a highly dynamic load profile for the main drives \cite{Kiselev+24}. 
Although the average motion-related power demand is significantly lower than the short-term peaks, the drive system can impose large transient loads on the observatory power infrastructure during demanding scan or slew manoeuvres. 
If dimensioned conventionally, such peaks would strongly influence the sizing of the off-grid electrical infrastructure, including converters, storage, and backup systems. 
Reducing these peaks is therefore important not only for energy efficiency, but also for limiting the size, cost, and environmental footprint of the renewable power system.

To this aim, the AtLAST drive architecture includes an energy recovery system (ERS) for both the azimuth and elevation drives.
Rather than dissipating kinetic energy as heat waste in braking resistors, kinetic energy is converted back into electrical energy by the motors during deceleration, where the ERS recovers most of it.
The ERS is analogous to the braking systems in hybrid/electric cars, but instead of using batteries, it is based on a supercapacitor energy buffer connected to the drive DC-link through a bidirectional DC/DC converter stage \cite{Kiselev+24}. This decouples the supercapacitor storage voltage from the DC-link voltage and enables controlled bidirectional energy exchange with the drive system.
The regenerated braking energy is then reused during subsequent acceleration phases, thereby improving the efficiency of telescope motion and shaving short high-power peaks seen by the renewable power infrastructure. 

For AtLAST, the ERS concept was first evaluated using representative fast-scanning motion profiles. 
A worst-case Lissajous daisy scan reaching the required maximum velocities and accelerations of the azimuth and elevation axes was used to derive the first-order system sizing \cite{Kiselev+24}. 
For this motion profile, the recoverable energy to be handled by the storage system is $\sim1.9$~MJ for both axes combined, setting the scale for the supercapacitor buffer and the bidirectional DC/DC converter stage \cite{Kiselev+24}. 
Dynamic simulations of the drive-side energy flow show that, for the considered representative AtLAST scan, the proposed system reduces the RMS grid-power demand of the main drives by about 56\% and reduces the drive-related grid-power peaks by about 80\% \cite{Kiselev+24}. 
The ERS therefore acts both as an energy-efficiency measure and as a peak-load mitigation element for the power infrastructure.

The ERS topology developed in the context of AtLAST has recently been implemented for the CTA South Large Size Telescope (LST) drive system \cite{Kiselev+26_CTA_ERS}. 
Although the system ratings are adapted to the specific motion requirements of each telescope, the underlying technical solution is the same. 
The two applications therefore demonstrate a key strength of the ERS concept: different operational challenges can be addressed by the same system. 
For the CTA South LST, the ERS mitigates rare but very high transient power demands during fast repositioning, while for AtLAST it supports high-duty-cycle fast-scanning operation and peak-load mitigation within an off-grid renewable power system. 
Performance tests with real hardware have successfully validated the CTA South LST implementation and moved the ERS concept beyond simulation \cite{Kiselev+26_CTA_ERS}. 
The CTA South LST implementation is therefore an important technology pathfinder for AtLAST, substantially increasing the TRL of the ERS concept for large-telescope drive systems. 
For AtLAST, the next steps will adapt these validated principles to the final drive architecture, including the common DC-link strategy between axes, the storage voltage window, converter modularity, supervisory control interfaces, and the operational constraints of the renewable off-grid power system.
The ERS will thus form an integral part of the sustainable observatory concept: rather than treating telescope motion as an unavoidable transient load, AtLAST will reuse part of its own braking energy and reduce the electrical infrastructure required to support fast-scanning observations.

\subsection{Main Axis Design}\label{sec:MAD}
The astronomical observation scenarios studied towards the end of AtLAST1 showed an increased emphasis on scanning compared to what was expected at the beginning of the design study, which led to a rise in load cycles along the main axes of the telescope structure. Within AtLAST2, we are paying particular attention to the mechanical architecture of the telescope support structure, especially with regard to load distribution, maintainability, and lifecycle considerations.  
The aforementioned conceptual design review (Section \ref{sec:CoDR}) indicated that the system may be exposed to load cycles approaching the endurance limit, suggesting that these conditions should be considered more explicitly in the design.
To address this, a comprehensive redesign of the bearing system is ongoing, with the aim of ensuring sufficient service life while enabling an effective maintenance concept that supports inspection and component replacement. A key requirement is also to harmonize the bearing principles of the elevation and azimuth axes as much as possible in order to simplify the maintenance process.

As part of this ongoing redesign, the number of bogies is being significantly increased to reduce the load per wheel. These so-called bogies are currently designed modular wheel assemblies that are hydraulically supported to achieve the most uniform load distribution possible among the individual units. Each bogie consists of two wheel sets, with each set comprising two wheels. This improved load distribution enhances fatigue performance and allows maintenance intervals to be maintained within acceptable operational limits. This evaluation is carried out with consideration of fatigue effects, including conditions beyond the endurance limit. To accommodate the increasing number of bogies, a comprehensive revision of the azimuthal structure is currently underway, simultaneously addressing all requirements necessary to facilitate the removal and replacement of bogies and rail segments within both the elevation and the azimuth bearing system.

\section{INSTRUMENTATION}\label{sec:inst}


As noted previously here, AtLAST will be able to host multiple instruments in six nominal instrument bays within a receiver cabin volume that is unprecedented in sub-mm astronomy. 
The instrument concepts were briefly outlined in AtLAST Memo \#3 \cite{AtLAST_memo_3}, and include: (1) an eight-band multi-chroic survey camera/polarimeter that could follow a similar approach as Simons Observatory\cite{Zhu2021}, scaling to $4-8\times$ larger and using the optic corrections adapted for AtLAST's FoV\cite{Gallardo+24}, (2) direct-detection spectroscopic integral field units (IFUs) featuring many spatial elements (beams, feeds) on sky and delivering spectral resolutions $R \equiv \nu / \delta \nu \approx 100-1000$ \cite{Kovacs2025, Endo+26_TIFUUN}, (3) heterodyne focal plane arrays including polarimetry and delivering $\sim 1$~km/s spectral resolutions both for sparse bright lines and for more crowded line forests,\cite{Groppi2019} and (4) multi-frequency single beam receivers\cite{Akiyama2023} that will enable participation of AtLAST in next generation very long baseline interferometry (VLBI), next generation Event Horizon Telescope (ngEHT), and, potentially, connected baseline campaigns with ALMA.  While the precise bands and numbers of elements on sky (for the multibeam instruments) will depend on the first light priorities for AtLAST, the majority of the science cases require frequencies across AtLAST's full range \cite{Booth+24_SPIE}.  
An upcoming work from the AtLAST instrumentation team (Mroczkowski et al.\ in prep) will detail the first generation instrumentation concepts for AtLAST, including technologies and their TRLs, sensitivities, spectral resolutions, and scalings such as costs and power usage.

The science demand for a multi-chroic camera for solar observations is also compelling\cite{2024ORE.....4..140W} and merits additional consideration due to the exceptional conditions under which it will operate. 
Simulations with the \texttt{maria} code demonstrate that a first-generation solar instrument could deliver full-disk, full-polarisation observations of the Sun in 4–8 simultaneous frequency bands with a cadence below 1\,min\cite{Kirkaune+25}. 
This first-generation solar instrument will require $\approx$50,000 pixels at the highest frequency to achieve instantaneous Nyquist sampling. A second-generation instrument with several hundred thousand pixels could reduce the cadence to only a few seconds. Among other applications, this capability would enable the detection and monitoring of solar flares anywhere on the solar disk -- an unprecedented capability with major implications for both fundamental solar and stellar physics and operational space-weather forecasting.
It is unclear at this stage if the aforementioned multi-chroic survey camera (case \#1 in the previous paragraph) could also perform these solar observations with the inclusion of a neutral density filter, or if a dedicated receiver should be developed and deployed for solar science.



In estimating instrument sensitivities for AtLAST, we rely in part on the significant upgrades to the AtLAST sensitivity calculator as part of the design consolidation phase. 
The latest version of the code can be found on GitHub or used online\footnote{\href{https://github.com/ukatc/AtLAST\_sensitivity\_calculator}{https://github.com/ukatc/AtLAST\_sensitivity\_calculator}, \href{https://www.atlast.uio.no/sensitivity-calculator/}{https://www.atlast.uio.no/sensitivity-calculator/}.}. 
The codebase was refactored into a modular architecture that separates user inputs, telescope and environmental parameters, derived quantities, and instrument definitions into clearly defined components. The new code consists of a central parameter management layer to coordinate these components. This allows the calculator to dynamically select appropriate instrument models based on user input observing frequencies and bandwidths, while still allowing user control when desired. 
The code currently embeds heterodyne spectrometers (case \#2 or \#4 above), a direct-detection continuum cameras (case \#1), and a direct-detection IFU (case \#3) as examples. 
New instrument modules can be added by users without having to modify the calculation engine. This lowers the barrier for science users and instrument teams to explore alternative designs, compare performance across instrument classes, and better connect observing strategies to science goals. 

For users that want to create mock observations of large AtLAST maps, similar to those shown in Figures~ \ref{fig:CO21_localgalaxy} \& ~\ref{fig:CI10_localgalaxy}, the \texttt{maria}\cite{vanMarrewijk+24} and \texttt{gateau}\cite{Moerman2026} tools are well suited for this task.


\section{SITE TESTING AND CHARACTERIZATION}\label{sec:site}
\subsection{Site selection}
\begin{figure}[tbp]
	\centering
	\includegraphics[clip=true,trim=0.cm 3.3cm 0.cm 2cm, width=\textwidth]{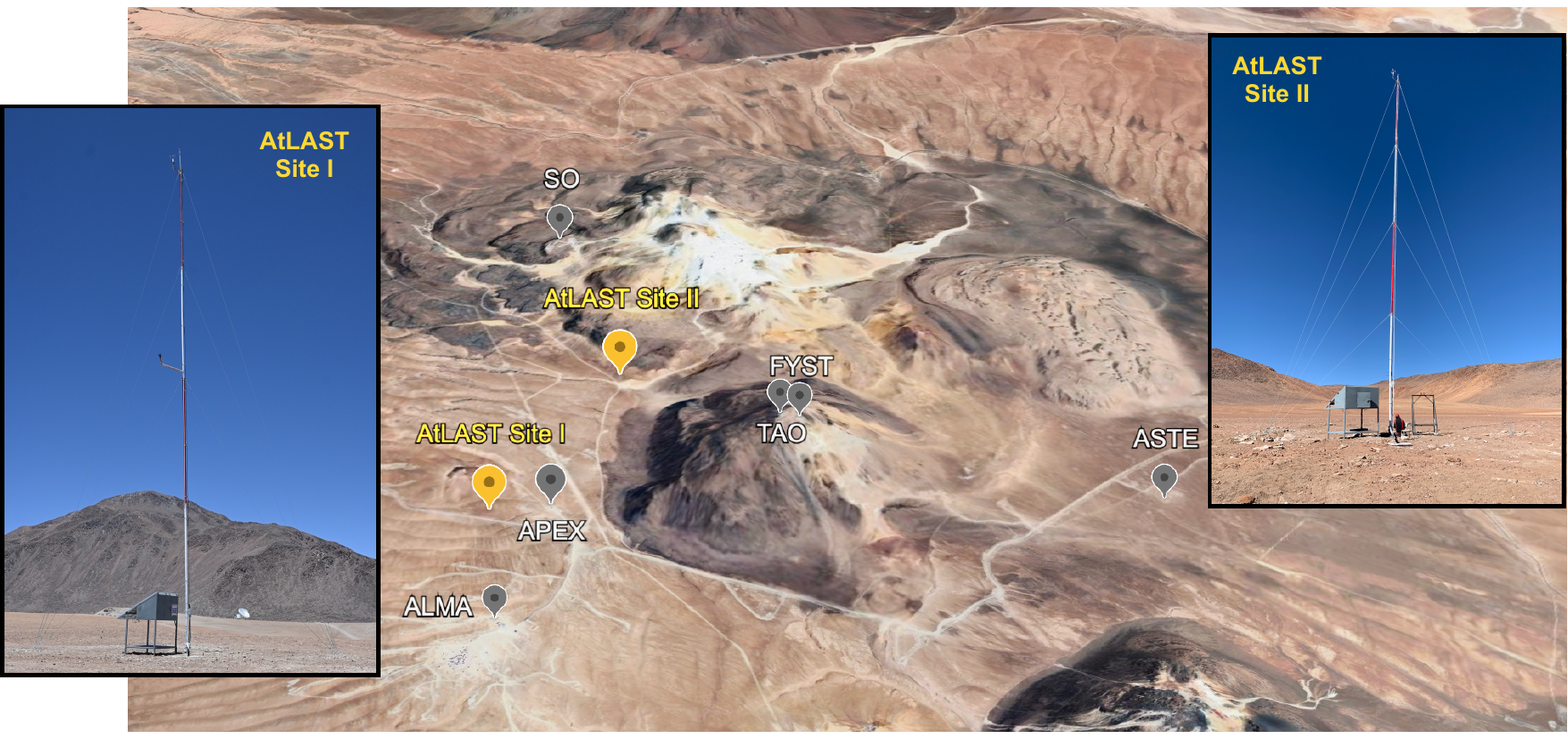}
	\caption{3D map of the Chajnantor Plateau and surrounding mountains showing the location of the two prospective AtLAST sites and other telescopes in the area. The insets display photos of the identical 24~m high AtLAST weather towers that have been collecting wind data simultaneously on both sites since 2023. The towers are equipped with sonic anemometers positioned at different heights. }
	\label{fig:sites_map}
\end{figure}

The selection criteria for the AtLAST site were defined during AtLAST1\cite{AtLAST_memo_9}. 
These include: excellent atmospheric transparency out to 950\,GHz, appropriate legal conditions to allow the construction of a large telescope, good accessibility and connectivity, a sufficiently large area without interference of nearby telescopes (e.g.\ shadowing of remote ALMA antenna pads), low snow coverage, minimal exposure to lightning strikes, low wind speed, and good atmospheric stability to minimize anomalous refraction. Based on the above, two potential sites were selected (see Fig.~\ref{fig:sites_map}): site I inside the ALMA concession and site II within the Atacama Astronomical Park, just outside the ALMA concession.

\subsection{Wind measurements}\label{subsec:wind}

One critical piece of missing information to down-select between the two prospective sites was the availability of high time-resolution data out to the $\sim$60~m height of the AtLAST telescope structure. To obtain these data, we constructed two identical weather towers of 24~m height on either of the sites (see Fig.~\ref{fig:sites_map}), each having multiple anemometers which sample the wind speed up to 20\,Hz time resolution at heights up to 24~m. By including anemometers at different heights, we can determine the surface roughness parameter and friction velocity, allowing to extrapolate the wind speed up to the full height of the telescope structure. These wind data are also informing the current telescope design consolidation work described in Sect.~\ref{sec:soap},  and have been used as input in the simulation-based dynamic pointing analysis presented by Kiselev et al.\cite{Kiselev+26_PE}. The site characterisation campaign will continue at least until the end of AtLAST2, allowing to sample both diurnal and seasonal variations. 

A preliminary analysis from the first 1.5 years of operations revealed substantially lower wind speeds on site~II, especially at nighttime (0-11 UTC) during the winter season (Apr-Sep), mostly because the prevailing winds from the W and NW are shielded by the nearby Cerro Toco mountain.
We estimated the number of hours per year that fulfill the following conditions: a) PWV$<$0.7\,mm and wind speed below 6\,m/s; and b) PWV$<$5\,mm and wind speed 12\,m/s. For this, we used
1~h average for wind speed measurements at 24~m, and PWV data obtained from the weather database of APEX for the same period of study.
The total estimated number of hours in a year in which conditions a) or b) are met is higher for site II than for site I. These amount to 999\,h for condition {\it a} and 4240\,h for condition {\it b} on site II, and 553\,h and 3437\,h respectively on site I.

\subsection{Phase measurements}\label{subsec:phase}

An additional parameter that needs to be determined to operate a 50~m single dish telescope is the anomalous refraction, which cannot be determined with anemometers of radiometers. Efforts are underway to characterize atmospheric turbulence over the envisaged 50~m baseline of the telescope at both proposed observation sites. This characterization will rely on atmospheric phase interferometry using Ku-band beacon signals transmitted from broadcast satellites, received by two off-the-shelf satellite television low-noise block downconverters sharing a common reference source. This approach was used successfully in site testing for the LMT \cite{hiriart2002}, and continues to be in use at the NOEMA Observatory \cite{Mahieu2019} where baseline RMS phase error of below 1\textdegree{} are reported routinely. 

\section{RENEWABLE ENERGY SYSTEM}\label{sec:renewable}



\subsection{Metal hydride hydrogen storage systems for the AtLAST infrastructure}\label{sec:MHs}

A bespoke renewable energy system set-up, consisting of photovoltaics (PVs) paired with a hybrid storage system including batteries, hydrogen storage, and a backup diesel generators, can secure a highly reliable power supply to the off-grid AtLAST facility \cite{Viole+23, Viole+24a, Viole+24b}. This also represents the most cost-effective solution in the 2030s and reduces the telescope's power-side carbon footprint by 95\% compared to the business-as-usual scenario.
While the batteries being considered are based on commercial technologies, in AtLAST2 we are conducting original research focused on
developing and testing tailored metal
hydrides intermetallic alloys for storage tanks and compressors.
These materials are optimised for AtLAST in terms of performance, cost-effectiveness, and environmental sustainability. 

The metal-hydride alloys at the focus of this work are so-called high entropy alloys (HEAs), also referred to as multicomponent alloys. HEAs consist of five or more principal elements, typically in concentrations ranging from 5 to 35\%, and have recently attracted significant attention for hydrogen storage applications due to their tuneable chemical and physical properties \cite{wang2021high}.
By varying the elemental composition, their thermodynamic and kinetic behaviour can be tailored for specific operational requirements, making them particularly suitable for specific case-studies such as the AtLAST infrastructure. 

Prior to material synthesis, we established specific design criteria: (i) the need of achieving hydrogen storage at moderate pressures (preferably below 40 bar) and near room temperature ($\sim296$~K); (ii) maintaining the desorption plateau pressure not lower than $\sim5$~bar to ensure hydrogen release under near-atmospheric pressure conditions, and (iii) avoiding the need for complex activation procedures in order to reduce system cost and enhance practical applicability. We selected materials based on their cost-effectiveness and sustainability, by prioritizing abundant, industrially-common elements and minimizing scarce or expensive ones (e.g., rare-earths). This also improves practical feasibility for large-scale applications. We adopted a compositional strategy in which a high fraction of non-hydride-forming elements ($\sim60–65$\%) was combined with hydride-forming elements ($\sim35–40$\%), enabling optimization of both hydrogen storage capacity and equilibrium pressure.

\begin{figure}
    \centering
    \includegraphics[clip=true,trim=0.cm 0.3cm 0.cm 0.3cm, width=0.45\linewidth]{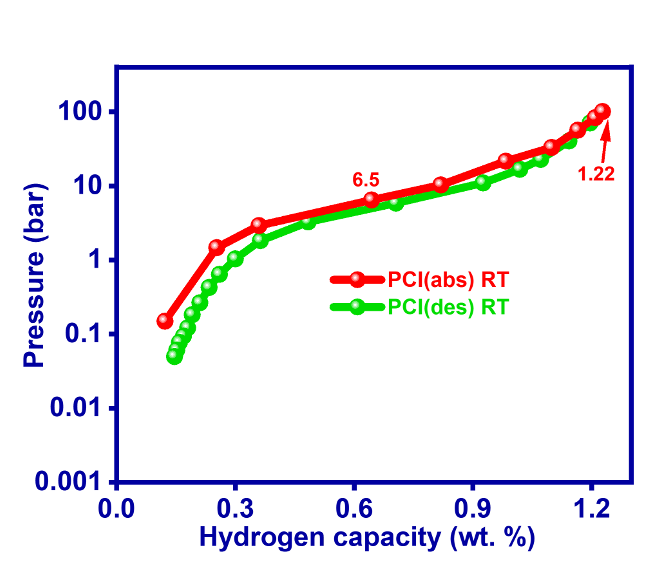}
    \caption{Pressure composition isotherms for hydrogen absorption and desorption at room temperature for the developed alloy within the Ti–Zr–Al–Cr–Fe HEA system, demonstrating reversible hydrogen storage without prior activation.}
    \label{fig:RES_figure}
\end{figure}

We carried out thermodynamic and phase stability predictions using the CALPHAD method based on the Thermo-Calc software with the TCHEA7 database \cite{bosi2023empirical} to select candidate compositions. The calculations targeted the formation of the C14 Laves phase (hexagonal crystal structure), which is known for its favourable hydrogen storage characteristics, including good cycling stability and moderate storage capacity. Based on the computational analyses, 
we synthesised 
representative compositions within the Ti–Zr–Al–Cr–Fe HEA system using electric arc melting. 
Structural and microstructural characterization carried out using X-ray diffraction and scanning electron microscopy confirmed the formation of the targeted phase. Pressure composition isotherm measurements at room temperature, shown in Fig.~\ref{fig:RES_figure}, demonstrate that hydrogen absorption and desorption occur readily without any prior activation process. 

One of the representative optimized composition exhibits a hydrogen storage capacity of approximately 1.22 wt.\%, indicating promising performance. The plateau pressure of $\sim6.5$~bar lies within a practical operating range, reflecting favourable thermodynamics for real-world applications, where hydrogen can be reversibly stored and released under moderate pressure conditions. The material further demonstrates reasonable cyclic stability over repeated absorption/desorption cycles. 
Future steps will consist in scaling up the materials, integrating them into prototype storage tanks and compression units, and validation at the system level. Pilot-scale testing of the hydride-based storage stack is planned through collaboration with the industrial partner GRZ\footnote{\href{https://www.grz-technologies.com/}{https://www.grz-technologies.com/}}. The prototype system is expected to provide a storage capacity of approximately 3~kg of hydrogen, and its performance will be evaluated under realistic environmental conditions, including potential deployment in the Atacama Desert near the AtLAST candidate sites.

\subsection{Community engagement and energy justice}\label{sec:socialjust}

Defining “sustainability” is challenging as the term often carries socio-economic assumptions, which are often associated with high-tech innovation, international trade, and eco-efficiency, grounded in a materialist, individualist and positivist worldview\cite{Dryzek2022}.  
It can therefore reflect an ethnocentric perspective that overlooks other worldviews, including those from the Global South \cite{velasco2022challenging} and, more specifically, from northern Chile, where AtLAST will be located.
From the outset, a core goal of AtLAST has been to integrate renewable energy into its design and to work with communities in the Chajnantor area to address the social justice implications of the project.

In AtLAST1 we analysed scenarios in which AtLAST’s energy system would also supply part of the electricity demand of San Pedro de Atacama, a nearby off-grid town that faces persistent power reliability and affordability challenges \cite{valenzuela2024renewable}\cite{velasco181}. These scenarios were discussed in stakeholder workshops and ranked according to expressed preferences.
In AtLAST2 we are focusing on local and regional understandings of sustainability, with the aim of aligning the telescope design and implementation  with these perspectives. 

We conducted interviews and focus groups with stakeholders at national (ministries, industry, universities), regional (regional government, research institutes), and local (municipalities, communities, and Indigenous communities) levels. To move beyond a narrowly modernist framing, we focused in particular on how people’s well being is affected by new research and economic infrastructures, and how these experiences shape their understandings of sustainability.
Preliminary findings suggest broad support for embedding sustainability considerations into AtLAST from this early stage. Stakeholders emphasised that this approach differs from previous scientific projects in the region and welcomed the change. At the same time, the AtLAST site is located next to mining industries and large-scale renewable energy developments, which bring economical benefits in terms of jobs and local revenues, but also generate conflicts over land use and recognition of Indigenous rights. Concerns were also raised about the unreliable regional energy infrastructure, the dominant role of private energy distributors, and the difficulty of engaging these actors in a project that aims to address energy accessibility and affordability. Participants also stressed the importance of revisiting sustainability questions not only during design, but also throughout implementation and operation.

Integrating a renewable energy system is essential to reducing emissions and enhancing the overall sustainability of AtLAST. Our preliminary results highlight the need to acknowledge local understandings of sustainability and wellbeing, and to move towards a human-centred energy system for AtLAST, one that supports climate mitigation while contributing to the wellbeing of neighbouring communities.

\subsection{Future work}

Operating AtLAST as an off-grid facility powered mainly by renewables requires a supervisory control system that coordinates, in real time, photovoltaic generation, battery and hydrogen storage, and backup diesel generators, in response to highly variable telescope demand. As the microgrid is isolated and dominated by power-electronic converters, it lacks the intrinsic stability that large interconnected grids provide.
Advanced control is therefore essential to ensure reliable, high-quality power supply.
The control framework under development for AtLAST combines forecasts of solar energy generation and telescope
load with predictive optimisation \cite{drgovna2020all} \cite{sultana2017review}, to schedule energy flows between generation, storage and backup units in advance, thereby   
improving reliability, extending the lifetime of storage assets, and minimising fuel consumption.

The supervisory control strategy under development will be based on Model Predictive Control (MPC), a family of control approaches that exploit a mathematical model of the system,
subject to operational constraints, to derive control actions through the minimisation of predefined cost functions. Within this framework, control decisions will be obtained by predicting
the future evolution of the system over a finite horizon and selecting the sequence of inputs that
optimises the chosen performance criteria. 
The optimisation layer will be responsible for determining the sequence of control actions
that best satisfies the operational objectives while respecting the physical and technical limits of
the system. In the context of the AtLAST hybrid energy system, these decisions will include the
scheduling of battery charging and discharging, the operation of the hydrogen production and
consumption chain, and the activation of backup generation units. By integrating forecasts of photovoltaic generation and telescope demand, the controller will anticipate future variations in energy availability
and load, and schedule energy flows among the different units in advance. This predictive
capability is expected to enhance the operational efficiency of the system, contribute to the
preservation of the storage assets, and ensure a reliable power supply to the facility, in line with the broader sustainability objectives pursued within AtLAST2.

\section{OPERATIONS}\label{sec:operations}

The operational modus of AtLAST is expected to evolve significantly throughout the facility lifecycle, transitioning from a predominantly on-site mode during construction and early phases, to a mature, remote distributed operational mode. This evolution is driven by the need to optimise operational efficiency, scientific output, and long-term sustainability, while also addressing inclusivity as well as economical and geopolitical considerations. 

Construction and Assembly, Integration, and Verification (AIV) activities 
will require substantial on-site presence 
while Commissioning will introduce a hybrid operational model in which telescope characterisation, calibration, and safety-critical activities continue to depend on in-site scientific and engineering staff. 
Through Science Verification (SV) and the transition to routine operations, a remote observing infrastructure, automated monitoring and scheduling, and distributed data-processing systems will progressively reduce the need for permanent on-site presence. The latter will be limited mainly to maintenance, safety oversight, and rapid technical intervention. Commissioning activities are expected to continue beyond the start of full operations, due to the regular integration of new instruments. While routine science operations can increasingly rely on remote execution and automation, instrument commissioning and upgrade activities will require periodic on-site scientific and technical campaigns. 

\subsection{AI-assisted operations}
The operations concepts for AtLAST are likely to rely heavily on Artificial Intelligence (AI)-assisted scheduling and decision support to maximise the scientific output of the facility in a remote and highly dynamic environment. At (sub-)mm wavelengths, atmospheric conditions can change rapidly and observing priorities may evolve on timescales of minutes to hours. At the same time, the enormous FoV of AtLAST in combination with parallel multi-wavelength observational capabilities opens the possibility to cover several scientific targets in a single observation. AI-driven dynamic scheduling systems could continuously evaluate weather forecasts, atmospheric transparency, instrument readiness, calibration status and scientific priorities, to optimise the scientific queue in real time. Such systems may also support coordinated campaigns with other observatories as well as rapid responses to transient alerts, enabling AtLAST to efficiently balance e.g.\ long-term survey programmes with time-critical observations, while minimising operational overheads.


AI methods are also expected to play a major role in predictive maintenance for facilities of the scale and complexity of AtLAST. By learning from the continuous monitoring of all systems involved in the telescope operations, AI tools may be able to identify subtle trends associated with component degradation or impeding failures before they affect observations. In addition to reducing downtime, such approaches could support condition-based maintenance planning, optimise spare parts logistics and improve long-term operational sustainability at remote, high-altitude sites like the one envisaged for AtLAST. While these concepts are currently exploratory in nature,  they represent promising directions for future efficient and resilient observatory operations. 

\subsection{Software architecture}

AtLAST is defining its high-level software architecture by assessing the processes and requirements of all observatory stakeholders. 
This approach ensures the system supports both the traditional proposal-to-observation workflow and emerging needs such as the real-time operational monitoring, automated data-driven observations scheduling and condition-based maintenance. By considering these diverse functions we aim to develop a system that meets the observatory's high-level requirements, while adhering to AtLAST's guiding principles: safety, high-performance, automation, sustainability, transparency, community involvement and inclusion.

The architecture is centred on the Science Archive, which serves as the single source of truth for all operational and scientific data. At the same time, it is designed to adapt to a distributed operations model, with multiple control centres coordinating telescope operations. Given the high expected  data rates, we adopt the modern paradigm of bringing users to the data through centralised analysis platforms and remote access tools, rather than distributing large datasets. This approach improves efficiency, scalability, and equitable access to data for the astronomical community. 
In developing this architecture, we draw on valuable lessons from a number of successful observatories, including APEX\cite{2018SPIE10704E..1VK} and ALMA \cite{2004SPIE.5496..190S}. We also incorporate emerging trends from newer facilities, like in the adoption of event-driven workflows, by leveraging cloud-native technologies, or prioritising open standards for interoperability. 

\subsection{Control system: APEX as a Pathfinder}

AtLAST expects to host a wide variety of frontends and backends throughout the observatory's lifetime. This will be similar to how the APEX telescope operates, which serves as a valuable frame of reference. To accommodate these instruments flexibly while avoiding to re-invent the ``interface wheel'' for each device, the APEX Control System (APECS) \cite{muders06} has adopted the concept of generic interface classes for a kind of device (e.g.\ heterodyne or continuum frontends, spectrometers, etc.)\footnote{\href{https://www3.mpifr-bonn.mpg.de/staff/dmuders/APEX/CORBA/APEX-MPI-ICD-0004-R4\_6.pdf}{https://www3.mpifr-bonn.mpg.de/staff/dmuders/APEX/CORBA/APEX-MPI-ICD-0004-R4\_6.pdf}}. Any new device of a given class must implement the same interface as all others so that it can be simply added as a new named object in APECS. Handling it for observing setups and configuration is thus homogenized. This also complies with the object-oriented design principle of ``separation of concerns'', as internal details of the particular instrument are not needed at the telescope control system (TCS) level. These generic interfaces have served APEX well for more than 20 years; however, anticipating that AtLAST will host several multi-pixel, multi-colour widefield instruments, identifying areas of improvement to accommodate emerging instrumentation technology becomes necessary. With this in mind, we have been conducting interviews with various instrument builders from MPIfR (for APEX), GBT, and SRT to collect shortcomings of the current interfaces and ideas for improvements.
Overall, there is a general desire for the instruments to convey more information about their internal status to the telescope control system, in particular if this would degrade the quality of observations, and thus trigger repair or maintenance actions. Currently, the APECS interface offers a high-level state machine, but with no feedback on the reason, which can create confusion for the operators and observers in case of error and failure. This should be a key consideration when designing future interface for AtLAST. In addition, the valid ranges for important monitoring and control points (e.g., cryo temperatures, LO state, pixel yield, etc.) should be fetched automatically from the embedded systems to maximize flexibility. 
In terms of raising errors and alarms, the experience at APEX shows that unregulated use of log messages can overload the telescope control system (e.g., repeating a message every 100 milliseconds if an embedded system believes that it is in a very critical condition). The interface layer, hence, needs to implement limitations on log message rates and possibly message aggregation based on repeated texts. 

While the initial APEX instruments were relatively simple with a clear separation into a frontend and a backend part, the new generation of devices is much more complex, in particular for array receivers/continuum cameras with hundreds or thousands of pixels, which are going to be the default at AtLAST with its large FoV. To adjust all pixels to optimal conditions, often an internal measurement of actual signals is needed ahead of the astronomical observations. Thus, controlling and data acquisition inside the instrument is required. This may have consequences for the interface toward the telescope control system for AtLAST.  In addition, instruments with multi-band setups have additional needs concerning the signal path from the telescope to the instruments. Mirrors and dichroics need to be set correctly, depending on which combination of bands is currently used. The existing interface in APECS has factored out the ``optics" devices as separate from the frontends, but often the joint knowledge is needed. This is another area of research that we will explore within AtLAST2. Some of the identified interface improvements are directly compatible with APECS, and will be tested at APEX. 

On the data acquisition side, APECS has also defined some data stream interface \footnote{\href{https://www3.mpifr-bonn.mpg.de/staff/dmuders/APEX/SCPI/APEX-MPI-ICD-0005-R1\_1.pdf}{https://www3.mpifr-bonn.mpg.de/staff/dmuders/APEX/SCPI/APEX-MPI-ICD-0005-R1\_1.pdf}},
which was optimized for heterodyne instruments and small continuum cameras. The advent of large cameras like AMKID\cite{reyes26} showed limitations in the internal format, which led to large overheads, in particular when using short dump times of just a few 10 ms. This provides an opportunity to explore new formats suitable for future wide-field multi-pixel instruments on AtLAST. 

\section{CONCLUSIONS}
AtLAST will be the highest-throughput single-dish (sub-)mm telescope yet; coupled with its massive collecting area, it will provide a mapping speed roughly two orders in magnitude faster than anything previously achieved. 
At the same time, transitioning to AtLAST requires scaling 
current single-dish instrumentation by orders of magnitude. This leap leaves the door wide open for regular upgrades, each successively enabling more ambitious and transformational science. Beyond the challenges unique to astronomical facilities, AtLAST2 addresses issues that are of interest to most large research infrastructures striving for resilience in a rapidly evolving technical, environmental, and geopolitical landscape. These include the processing and storage of large datasets, remote and AI-assisted operations, preventive maintenance, reduced energy consumption, environmental sustainability, and social acceptance.
Ultimately, AtLAST aspires to be the first climate-neutral modern research infrastructure, offering a bold vision of sustainable, transformational astronomy, that will shape the future of scientific research facilities world-wide.

\acknowledgments 
 
This project has received funding from the European Union's Horizon Europe and Horizon 2020 research and innovation programmes under grant agreements No. 101188037 (AtLAST2) and No. 951815 (AtLAST). Views and opinions expressed are however those of the author(s) only and do not necessarily reflect those of the European Union or European Research Executive Agency. Neither the European Union nor the European Research Executive Agency can be held responsible for them. 
The authors thank all AtLAST team members and collaborators, who are much more numerous than the list of authors of this manuscript. We thank Silvia Piranomonte, Cristian Vignali, Paola Severgnini, and several collaborators on the AtLAST slack channel for very valuable feedback on Fig.~\ref{fig:mapping_speed}. 
The map in Figure~\ref{fig:sites_map} was created using Google Earth and it includes data from AirbusLandsat, CopernicusData SIO, NOAA, U.S. Navy, NGA, GEBCOMaxar TechnologiesImagery from the dates 14 Dec 2015 - 12 Jan 2025.
TM acknowledges support from the Agencia Estatal de Investigaci\'on (AEI) and the Ministerio de Ciencia, Innovaci\'on y Universidades (MICIU) Grant ATRAE2024-154740 funded by MICIU/AEI//10.13039/501100011033.
TM, FK, and SR are also partly supported by the Spanish program Unidad de Excelencia María de Maeztu CEX2020-001058-M, financed by MCIN/AEI//10.13039/501100011033, and by the MaX-CSIC Excellence Award MaX4-SOMMA-ICE.
FMM acknowledges support from Spanish national project PID2024-157374OB-I00, funded by MICIU/AEI/ 10.13039/501100011033/ FEDER, UE. 
FK acknowledges support from Spanish national project PID2023-149918NB-I00, funded by MICIU/ AEI/ 10.13039/ 501100011033/ FEDER, EU.
SW and MK are supported by the Research Council of Norway through the Centres of Excellence scheme, project number 262622 (“Rosseland Centre for Solar Physics”). T. Stander's contribution is co-funded by the South African Radio Astronomy Observatory (SARAO) under Grant UID97929.
    
\bibliography{SPIE2026_AtLAST} 

@ARTICLE{Mroczkowski+25,
	author = {{Mroczkowski}, Tony and {Gallardo}, Patricio A. and {Timpe}, Martin and {Kiselev}, Aleksej and {Groh}, Manuel and {Kaercher}, Hans and {Reichert}, Matthias and {Cicone}, Claudia and {Puddu}, Roberto and {Dubois-dit-Bonclaude}, Pierre and {Bok}, Daniel and {Dahl}, Erik and {Macintosh}, Mike and {Dicker}, Simon and {Viole}, Isabelle and {Sartori}, Sabrina and {Valenzuela Venegas}, Guillermo Andr{\'e}s and {Zeyringer}, Marianne and {Niemack}, Michael and {Poppi}, Sergio and {Olguin}, Rodrigo and {Hatziminaoglou}, Evanthia and {De Breuck}, Carlos and {Klaassen}, Pamela and {Montenegro-Montes}, Francisco Miguel and {Zimmerer}, Thomas},
	title = "{The conceptual design of the 50-meter Atacama Large Aperture Submillimeter Telescope (AtLAST)}",
	journal = {\aap},
	year = 2025,
	month = feb,
	volume = {694},
	eid = {A142},
	pages = {A142},
	doi = {10.1051/0004-6361/202449786},
	archivePrefix = {arXiv},
	eprint = {2402.18645},
	primaryClass = {astro-ph.IM},
	adsurl = {https://ui.adsabs.harvard.edu/abs/2025A&A...694A.142M}
}

@ARTICLE{reyes26,
       author = {{Reyes}, N. and {Weiss}, A. and {Yates}, S.~J.~C. and {Baryshev}, A.~M. and et al.},
        title = "{AMKID: A large KID-based camera at the APEX telescope}",
      journal = {\aap},
         year = 2026,
        month = mar,
       volume = {707},
          eid = {A294},
        pages = {A294},
          doi = {10.1051/0004-6361/202558596},
archivePrefix = {arXiv},
       eprint = {2512.14905},
 primaryClass = {astro-ph.IM},
       adsurl = {https://ui.adsabs.harvard.edu/abs/2026A&A...707A.294R}
}

@inproceedings{Klaassen+20,
	author = {Pamela D. Klaassen and Tony K. Mroczkowski and Claudia Cicone and Evanthia Hatziminaoglou and Sabrina Sartori and Carlos De Breuck and Sean Bryan and Simon R. Dicker and Carlos Duran and Chris Groppi and Hans Kaercher and Ryohei Kawabe and Kotaro Kohno and James Geach},
	title = {{The Atacama Large Aperture Submillimeter Telescope (AtLAST)}},
	volume = {11445},
	booktitle = {Ground-based and Airborne Telescopes VIII},
	editor = {Heather K. Marshall and Jason Spyromilio and Tomonori Usuda},
	organization = {International Society for Optics and Photonics},
	publisher = {SPIE},
	pages = {114452F},
	year = {2020},
	doi = {10.1117/12.2561315},
	URL = {https://doi.org/10.1117/12.2561315}
}

@INPROCEEDINGS{Mroczkowski+23,
	author={Mroczkowski, Tony and Cicone, Claudia and Reichert, Matthias and Gallardo, Patricio and Kaercher, Hans and Hills, Richard and Bok, Daniel and Dahl, Erik and Dubois-Dit-Bonclaude, Pierre and Kiselev, Aleksej and Timpe, Martin and Zimmerer, Thomas and Dicker, Simon and Macintosh, Mike and Klaassen, Pamela and Niemack, Michael},
	booktitle={2023 XXXVth General Assembly and Scientific Symposium of the International Union of Radio Science (URSI GASS)}, 
	title={Progress in the Design of the Atacama Large Aperture Submillimeter Telescope}, 
	year={2023},
	volume={},
	number={},
	pages={1-4},
	doi={10.23919/URSIGASS57860.2023.10265372}}

@inproceedings{Reichert+24,
	author = {Matthias Reichert and Martin Timpe and Hans Kaercher and Tony Mroczkowski and Manuel Groh and Aleksej Kiselev and Claudia Cicone and Patricio Gallardo and Roberto Puddu and Pamela Klaasen},
	title = {{Technical requirements flow-down for the concept design of the novel 50-meter Atacama Large Aperture Submm Telescope (AtLAST)}},
	volume = {13094},
	booktitle = {Ground-based and Airborne Telescopes X},
	editor = {Heather K. Marshall and Jason Spyromilio and Tomonori Usuda},
	organization = {International Society for Optics and Photonics},
	publisher = {SPIE},
	pages = {130941U},
	year = {2024},
	doi = {10.1117/12.3018133},
archivePrefix = {arXiv},
    eprint = {2406.08611},
 primaryClass = {astro-ph.IM},
    adsurl = {https://ui.adsabs.harvard.edu/abs/2024SPIE13094E..1UR},
    URL = {https://doi.org/10.1117/12.3018133}
}

@inproceedings{Gallardo+24,
	author = {Patricio A. Gallardo and Roberto Puddu and Tony Mroczkowski and Martin Timpe and Pierre Dubois-dit-Bonclaude and Manuel Groh and Matthias Reichert and Claudia Cicone and Hans J. Kaercher},
	title = {{The optical design concept for the Atacama Large Aperture Submillimeter Telescope (AtLAST)}},
	volume = {13094},
	booktitle = {Ground-based and Airborne Telescopes X},
	editor = {Heather K. Marshall and Jason Spyromilio and Tomonori Usuda},
	organization = {International Society for Optics and Photonics},
	publisher = {SPIE},
	pages = {1309428},
	year = {2024},
	doi = {10.1117/12.3020272},
    adsurl = {https://ui.adsabs.harvard.edu/abs/2024SPIE13094E..28G},
	URL = {https://doi.org/10.1117/12.3020272}
}

@inproceedings{Puddu+24,
	author = {Roberto Puddu and Patricio A. Gallardo and Tony Mroczkowski and Pierre Dubois-dit-Bonclaude and Manuel Groh and Aleksej Kiselev and Matthias Reichert and Martin Timpe and Claudia Cicone and Hans J. Kaercher and Rolando D{\"u}nner},
	title = {{A physical optics characterization of the beam shape and sidelobe levels for the Atacama Large Aperture Submillimeter Telescope (AtLAST)}},
	volume = {13094},
	booktitle = {Ground-based and Airborne Telescopes X},
	editor = {Heather K. Marshall and Jason Spyromilio and Tomonori Usuda},
	organization = {International Society for Optics and Photonics},
	publisher = {SPIE},
	pages = {130944S},
	year = {2024},
	doi = {10.1117/12.3020773},
	URL = {https://doi.org/10.1117/12.3020773}
}

@inproceedings{Booth+24_SPIE,
	author = {Mark Booth and Pamela Klaassen and Claudia Cicone and Tony Mroczkowski and Sven Wedemeyer and Kazunori Akiyama and Geoffrey Bower and Martin A. Cordiner and Luca Di Mascolo and Doug Johnstone and Eelco van Kampen and Minju M. Lee and Daizhong Liu and John Orlowski-Scherer and Am{\'e}lie Saintonge and Matthew Smith and Alexander E. Thelen},
	title = {{The key science drivers for the Atacama Large Aperture Submillimeter Telescope (AtLAST)}},
	volume = {13102},
	booktitle = {Millimeter, Submillimeter, and Far-Infrared Detectors and Instrumentation for Astronomy XII},
	editor = {Jonas Zmuidzinas and Jian-Rong Gao},
	organization = {International Society for Optics and Photonics},
	publisher = {SPIE},
	pages = {1310206},
	year = {2024},
	doi = {10.1117/12.3017058},
	URL = {https://doi.org/10.1117/12.3017058}
}

@inproceedings{Kiselev+24,
	author = {Aleksej Kiselev and Matthias Reichert and Tony Mroczkowski},
	title = {{Energy recovery system for large telescopes}},
	volume = {13094},
	booktitle = {Ground-based and Airborne Telescopes X},
	editor = {Heather K. Marshall and Jason Spyromilio and Tomonori Usuda},
	organization = {International Society for Optics and Photonics},
	publisher = {SPIE},
	pages = {130940E},
	year = {2024},
	doi = {10.1117/12.3017762},
	URL = {https://doi.org/10.1117/12.3017762}
}

@article{Viole+23,
	title = {A renewable power system for an off-grid sustainable telescope fueled by solar power, batteries and green hydrogen},
	journal = {Energy},
	volume = {282},
	pages = {128570},
	year = {2023},
	issn = {0360-5442},
	doi = {https://doi.org/10.1016/j.energy.2023.128570},
	url = {https://www.sciencedirect.com/science/article/pii/S0360544223019643},
	author = {Isabelle Viole and Guillermo Valenzuela-Venegas and Marianne Zeyringer and Sabrina Sartori}
}

@article{Viole+24a,
	author    = {Viole, Isabelle and Shen, Li and Ramirez Camargo, Luis and Zeyringer, Marianne and Sartori, Sabrina},
	title     = {Sustainable astronomy: A comparative life cycle assessment of off-grid hybrid energy systems to supply large telescopes},
	journal   = {The International Journal of Life Cycle Assessment},
	year      = {2024},
	volume    = {29},
	pages     = {1706--1726},
	month     = may,
	doi       = {10.1007/s11367-024-02288-9},
	url       = {https://link.springer.com/article/10.1007/s11367-024-02288-9},
	note      = {Open access}
}

@article{Viole+24b,
	title = {Integrated life cycle assessment in off-grid energy system design—Uncovering low hanging fruit for climate mitigation},
	journal = {Applied Energy},
	volume = {367},
	pages = {123334},
	year = {2024},
	issn = {0306-2619},
	doi = {https://doi.org/10.1016/j.apenergy.2024.123334},
	url = {https://www.sciencedirect.com/science/article/pii/S0306261924007177},
	author = {Isabelle Viole and Guillermo Valenzuela-Venegas and Sabrina Sartori and Marianne Zeyringer},
}

@ARTICLE{Schimek+24a,
	author = {{Schimek}, A. and {Decataldo}, D. and {Shen}, S. and {Cicone}, C. and {Baumschlager}, B. and {van Kampen}, E. and {Klaassen}, P. and {Madau}, P. and {Di Mascolo}, L. and {Mayer}, L. and {Montoya Arroyave}, I. and {Mroczkowski}, T. and {Warraich}, J.},
	title = "{High resolution modelling of [CII], [CI], [OIII], and CO line emission from the interstellar medium and circumgalactic medium of a star-forming galaxy at z {\ensuremath{\sim}} 6.5}",
	journal = {\aap},
	year = 2024,
	month = feb,
	volume = {682},
	eid = {A98},
	pages = {A98},
	doi = {10.1051/0004-6361/202346945},
	archivePrefix = {arXiv},
	eprint = {2306.00583},
	primaryClass = {astro-ph.GA},
	adsurl = {https://ui.adsabs.harvard.edu/abs/2024A&A...682A..98S}
}

@ARTICLE{Schimek+24b,
	author = {{Schimek}, A. and {Cicone}, C. and {Shen}, S. and {Decataldo}, D. and {Klaassen}, P. and {Mayer}, L.},
	title = "{Constraining the physical properties of gas in high-z galaxies with far-infrared and submillimetre line ratios}",
	journal = {\aap},
	year = 2024,
	month = jul,
	volume = {687},
	eid = {L10},
	pages = {L10},
	doi = {10.1051/0004-6361/202449903},
	archivePrefix = {arXiv},
	eprint = {2406.11559},
	primaryClass = {astro-ph.GA},
	adsurl = {https://ui.adsabs.harvard.edu/abs/2024A&A...687L..10S}
}

@ARTICLE{Kirkaune+25,
	author = {{Kirkaune}, Mats and {Wedemeyer}, Sven and {van Marrewijk}, Joshiwa and {Mroczkowski}, Tony and {Morris}, Thomas W.},
	title = "{Observing the Sun with the Atacama Large Aperture Submillimeter Telescope (AtLAST): Forecasting Full-disk Observations}",
	journal = {The Open Journal of Astrophysics},
	year = 2025,
	month = sep,
	volume = {8},
	eid = {129},
	pages = {129},
	doi = {10.33232/001c.144106},
	archivePrefix = {arXiv},
	eprint = {2505.13145},
	primaryClass = {astro-ph.SR},
	adsurl = {https://ui.adsabs.harvard.edu/abs/2025OJAp....8E.129K}
}

@article{vanMarrewijk+24,
	author    = {van Marrewijk, J. and Morris, T. W. and Mroczkowski, T. and Cicone, C. and Dicker, S. and Di Mascolo, L. and Haridas, S. K. and Orlowski-Scherer, J. and Rasia, E. and Romero, C. and W{\"u}rzinger, J.},
	title     = {\textit{maria}: A Novel Simulator for Forecasting (Sub-)mm Observations},
	journal   = {The Open Journal of Astrophysics},
	year      = {2024},
	volume    = {7},
	month     = dec,
	doi       = {10.33232/001c.127571},
	url       = {https://astro.theoj.org/article/127571-_maria_-a-novel-simulator-for-forecasting-sub-mm-observations}
}

@ARTICLE{Morris+22,
       author = {{Morris}, Thomas W. and {Bustos}, Ricardo and {Calabrese}, Erminia and {Choi}, Steve K. and {Duivenvoorden}, Adriaan J. and {Dunkley}, Jo and {D{\"u}nner}, Rolando and {Gallardo}, Patricio A. and {Hasselfield}, Matthew and {Hincks}, Adam D. and {Mroczkowski}, Tony and {Naess}, Sigurd and {Niemack}, Michael D. and {Page}, Lyman and {Partridge}, Bruce and {Salatino}, Maria and {Staggs}, Suzanne and {Treu}, Jesse and {Wollack}, Edward J. and {Xu}, Zhilei},
        title = "{The Atacama Cosmology Telescope: Modeling bulk atmospheric motion}",
      journal = {\prd},
         year = 2022,
        month = feb,
       volume = {105},
       number = {4},
          eid = {042004},
        pages = {042004},
          doi = {10.1103/PhysRevD.105.042004},
archivePrefix = {arXiv},
       eprint = {2111.01319},
 primaryClass = {astro-ph.IM},
       adsurl = {https://ui.adsabs.harvard.edu/abs/2022PhRvD.105d2004M}
}

@article{wang2021high,
  title={High-entropy alloys: emerging materials for advanced functional applications},
  author={Wang, Xin and Guo, Wei and Fu, Yongzhu},
  journal={Journal of Materials Chemistry A},
  volume={9},
  number={2},
  pages={663--701},
  year={2021},
  publisher={Royal Society of Chemistry}
}

@article{bosi2023empirical,
  title={Empirical and computational-based phase predictions of thermal sprayed high-entropy alloys},
  author={Bosi, Ecio and Meghwal, Ashok and Singh, Surinder and Munroe, Paul and Berndt, Christopher C and Ang, Andrew Siao Ming},
  journal={Journal of Thermal Spray Technology},
  volume={32},
  number={6},
  pages={1840--1855},
  year={2023},
  publisher={Springer}
}

@ARTICLE{Bonanomi+24,
       author = {{Bonanomi}, Francesca and {Hacar}, Alvaro and {Socci}, Andrea and {Petry}, Dirk and {Suri}, S{\"u}meyye},
        title = "{Emergence of high-mass stars in complex fiber networks (EMERGE). II. The need for data combination in ALMA observations}",
      journal = {\aap},
         year = 2024,
        month = aug,
       volume = {688},
          eid = {A30},
        pages = {A30},
          doi = {10.1051/0004-6361/202348920},
archivePrefix = {arXiv},
       eprint = {2405.09290},
 primaryClass = {astro-ph.GA},
       adsurl = {https://ui.adsabs.harvard.edu/abs/2024A&A...688A..30B}
}

@ARTICLE{Koljonen+25_ESOWP,
       author = {{Koljonen}, Karri and {Ricci}, Claudio and {Stanke}, Thomas and {Johnstone}, Doug and {Mohan}, Atul and {Montenegro-Montes}, Francisco and {Orlowski-Scherer}, John},
        title = "{Serendipitous and targeted mm/sub-mm transient searches with wide-FOV telescope}",
      journal = {arXiv e-prints},
         year = 2025,
        month = dec,
          eid = {arXiv:2512.14768},
        pages = {arXiv:2512.14768},
          doi = {10.48550/arXiv.2512.14768},
archivePrefix = {arXiv},
       eprint = {2512.14768},
 primaryClass = {astro-ph.IM},
       adsurl = {https://ui.adsabs.harvard.edu/abs/2025arXiv251214768K}
}

@ARTICLE{Wedemeyer+25_ESOWP,
       author = {{Wedemeyer}, Sven and {Poedts}, Stefaan and {Gun{\'a}r}, Stanislav and {Temmer}, Manuela and {Veronig}, Astrid and {Nakariakov}, Valery and {Kirkaune}, Mats and {Cicone}, Claudia and {White}, Stephen and {Magdaleni{\'c}}, Jasmina and {Braj{\v{s}}a}, Roman and {De Pontieu}, Bart and {Saberi}, Maryam and {Mohan}, Atul and {Sudar}, Davor and {Motorina}, Galina and {Lukicheva}, Maria and {Sim{\~o}es}, Paulo},
        title = "{Millimeter-Wavelength Observations of the Active Sun: Unveiling the Origins of Space Weather}",
      journal = {arXiv e-prints},
         year = 2025,
        month = dec,
          eid = {arXiv:2512.13813},
        pages = {arXiv:2512.13813},
          doi = {10.48550/arXiv.2512.13813},
archivePrefix = {arXiv},
       eprint = {2512.13813},
 primaryClass = {astro-ph.IM},
       adsurl = {https://ui.adsabs.harvard.edu/abs/2025arXiv251213813W}
}

@article{valenzuela2024renewable,
  title={A renewable and socially accepted energy system for astronomical telescopes},
  author={Valenzuela-Venegas, Guillermo and Lode, Maria Luisa and Viole, Isabelle and Felice, Alex and Martinez Alonso, Ander and Ramirez Camargo, Luis and Sartori, Sabrina and Zeyringer, Marianne},
  journal={Nature Sustainability},
  volume={7},
  number={12},
  pages={1642--1650},
  year={2024},
  publisher={Nature Publishing Group UK London}
}

@inproceedings{Thoms+26_WFDSensorPlacement,
  author    = {Thoms, Stefan and Timpe, Martin},
  title     = {Optimizing wavefront-deformation sensor placement for active radio-telescope surfaces},
  booktitle = {Proceedings of SPIE Astronomical Telescopes and Instrumentation},
  year      = {2026}
}

@inproceedings{Kiselev+26_CTA_ERS,
  author    = {Kiselev, Aleksej and Klapp, Fabian and Giannouloudis, Konstantinos and Petersen, Arne and Hitzmann, Hauke and Ak, Dennis},
  title     = {Toward Sustainable Telescope Motion: Supercapacitor-Based Energy Recovery for CTA South Large Size Telescopes},
  booktitle = {Proceedings of SPIE Astronomical Telescopes and Instrumentation},
  year      = {2026}
}

@inproceedings{Kiselev+26_PE,
  author    = {Kiselev, Aleksej and Timpe, Martin and Reichert, Matthias and Mroczkowski, Tony and Cicone, Claudia},
  title     = {Simulation-based dynamic pointing analysis of AtLAST under
wind loading and fast-scan conditions},
  booktitle = {Proceedings of SPIE Astronomical Telescopes and Instrumentation},
  year      = {2026}
}

@inproceedings{Endo+26_TIFUUN,
  author = {Akira Endo and Tom J. L. C. Bakx and Jochem J. A. Baselmans and Stefanie A. Brackenhoff and Bernhard R. Brandl and Dries Boleij and Martino Calvo and Shahab O. Dabironezare and Hans van der Does and Rei Enokiya and Tristan Oude Essink and Sho Fujisawa and Shinji Fujita and Enrico Garaldi and Wouter Gregoor and Masato Hagimoto and Davit Hakobyan and Angelina Harke-Hosemann and Robert Huiting and Shiro Ikeda and Kenichi Karatsu and Nick de Keijzer and Takumi Kojima and Alkistis Kyriakidou and Kotaro Kohno and Reinier M. J. Janssen and Louis H. Marting and Tomotake Matsumura and Cory Meijneke and Tetsuhiro Minamidani and Arend Moerman and Kana Moriwaki and Alessandro Monfardini and Kanako Narita and Yuri Nishimura and Erika Ogata and Leon G. G. Olde Scholtenhuis and Jim R. Piek and Matus Rybak and Kana Sakaguri and Seiichi Sakamoto and Aurora Simionescu and Nikita A. Soshnin and Tatsuya Takekoshi and Akio Taniguchi and David J. Thoen and Yoichi Tamura and Sten Vollebregt and Lingyu Wang and Paul P. van der Werf and Stephen J. C. Yates and Naoki Yoshida and Silvia Zhang},
  title     = {Development of TIFUUN: Terahertz Integral Field Units with Universal Nanotechnology},
  booktitle = {Proceedings of SPIE Astronomical Telescopes and Instrumentation},
  series = {Proceedings of SPIE},
  publisher = {SPIE},
  year      = {2026}
}

@techreport{Narayanan+20_VLAmemo,
  author  = {Narayanan, Desika and Jimenez-Andradez, Eric F.  and  Murphy,Eric and Li,  Qi  and  Rosero, Viviana},
  title   = {Imaging Cold Gas in High-Redshift galaxies
with the ngVLA: Memo Number 83},
  year    = {2020},
  institution = {VLA Memo Series},
  type    = {Technical Memo},
  number  = {VLA Memo number 83},
  url     = {https://library.nrao.edu/public/memos/ngvla/NGVLA_83.pdf},
  urldate = {2020-09-17}
}

@INPROCEEDINGS{2004SPIE.5496..190S,
       author = {{Schwarz}, Joseph and {Farris}, Allen and {Sommer}, Heiko},
        title = "{The ALMA software architecture}",
    booktitle = {Advanced Software, Control, and Communication Systems for Astronomy},
         year = 2004,
       editor = {{Lewis}, Hilton and {Raffi}, Gianni},
       series = {Society of Photo-Optical Instrumentation Engineers (SPIE) Conference Series},
       volume = {5496},
        month = sep,
        pages = {190-204},
          doi = {10.1117/12.551712},
       adsurl = {https://ui.adsabs.harvard.edu/abs/2004SPIE.5496..190S}
}

@INPROCEEDINGS{2018SPIE10704E..1VK,
       author = {{Klein}, Thomas and {Montenegro-Montes}, Francisco M. and {Ciechanowicz}, Miroslaw and {Agurto}, Claudio and et al.},
        title = "{APEX beyond 2016: the evolution of an experiment into an efficient and productive Submillimeter Wavelength Observatory}",
    booktitle = {Observatory Operations: Strategies, Processes, and Systems VII},
         year = 2018,
       series = {Society of Photo-Optical Instrumentation Engineers (SPIE) Conference Series},
       volume = {10704},
        month = jul,
          eid = {107041V},
        pages = {107041V},
          doi = {10.1117/12.2312687},
       adsurl = {https://ui.adsabs.harvard.edu/abs/2018SPIE10704E..1VK}
}

@incollection{velasco181,
  title={181 Astronomy and energy justice in the Atacama Desert},
  author={Velasco-Herrej{\'o}n, Paola and Viole, Isabelle and Valenzuela-Venegas, Guillermo and Sartori, Sabrina and Zeyringer, Marianne},
  booktitle={Energy Justice in Latin America: Reflections, Lessons and Critiques},
  pages={181--198},
  publisher={Routledge}
}

@article{velasco2022challenging,
  title={Challenging dominant sustainability worldviews on the energy transition: Lessons from Indigenous communities in Mexico and a plea for pluriversal technologies},
  author={Velasco-Herrej{\'o}n, Paola and Bauwens, Thomas and Friant, Martin Calisto},
  journal={World Development},
  volume={150},
  pages={105725},
  year={2022},
  publisher={Elsevier}
}

@article{drgovna2020all,
  title={All you need to know about model predictive control for buildings},
  author={Drgo{\v{n}}a, J{\'a}n and Arroyo, Javier and Figueroa, Iago Cupeiro and Blum, David and Arendt, Krzysztof and Kim, Donghun and Oll{\'e}, Enric Perarnau and Oravec, Juraj and Wetter, Michael and Vrabie, Draguna L and others},
  journal={Annual reviews in control},
  volume={50},
  pages={190--232},
  year={2020},
  publisher={Elsevier}
}

@article{sultana2017review,
  title={A review on state of art development of model predictive control for renewable energy applications},
  author={Sultana, W Razia and Sahoo, Sarat Kumar and Sukchai, Sukruedee and Yamuna, S and Venkatesh, D},
  journal={Renewable and sustainable energy reviews},
  volume={76},
  pages={391--406},
  year={2017},
  publisher={Elsevier}
}

@ARTICLE{Hiriart2002,
  title     = "Radio Seeing Monitor Interferometer",
  author    = "Hiriart, David and Valdez, Jorge and Zaca, Placido and Medina,
               Jos{\'e} L",
  journal   = "Publ. Astron. Soc. Pac.",
  publisher = "IOP Publishing",
  volume    =  114,
  number    =  800,
  pages     = "1150--1155",
  month     =  oct,
  year      =  2002
}

@INPROCEEDINGS{Mahieu2019,
       author = {{Mahieu}, S. and {Pissard}, B. and {Bremer}, M. and {Risacher}, C. and {Blundell}, R. and {Kimberk}, R. and {Test}, J.},
        title = "{Atmospheric Phase Monitoring Interferometer for the NOEMA Observatory}",
    booktitle = {Proceedings of the 30th International Symposium on Space Terahertz Technology},
         year = 2019,
        month = apr,
        pages = {149},
       adsurl = {https://ui.adsabs.harvard.edu/abs/2019stt..conf..149M}
}

@INPROCEEDINGS{Tamura2020,
       author = {{Tamura}, Yoichi and {Kawabe}, Ryohei and {Fukasaku}, Yuhei and {Kimura}, Kimihiro and {Ueda}, Tetsutaro and {Taniguchi}, Akio and {Okada}, Nozomi and {Ogawa}, Hideo and {Hashimoto}, Ikumi and {Minamidani}, Tetsuhiro and {Kawaguchi}, Noriyuki and {Kuno}, Nario and {Togami}, Yohei and {Hagimoto}, Masato and {Nakano}, Satoya and {Matsuda}, Keiichi and {Okumura}, Sachiko and {Nakamura}, Tomoko and {Kurita}, Mikio and {Takekoshi}, Tatsuya and {Oshima}, Tai and {Onishi}, Toshikazu and {Kohno}, Kotaro},
        title = "{Wavefront sensor for millimeter/submillimeter-wave adaptive optics based on aperture-plane interferometry}",
    booktitle = {Ground-based and Airborne Telescopes VIII},
         year = 2020,
       editor = {{Marshall}, Heather K. and {Spyromilio}, Jason and {Usuda}, Tomonori},
       series = {Society of Photo-Optical Instrumentation Engineers (SPIE) Conference Series},
       volume = {11445},
        month = dec,
          eid = {114451N},
        pages = {114451N},
          doi = {10.1117/12.2561885},
archivePrefix = {arXiv},
       eprint = {2102.09286},
 primaryClass = {astro-ph.IM},
       adsurl = {https://ui.adsabs.harvard.edu/abs/2020SPIE11445E..1NT}
}

@INPROCEEDINGS{Attoli2023,
       author = {{Attoli}, A. and {Poppi}, S. and {Buffa}, F. and {Serra}, G. and {Fara}, A. and {Marongiu}, P. and {Sanna}, G. and {Gaudiomonte}, F. and {Pili}, M. and {Pisanu}, T. and {Vargiu}, G.~P. and {Fierro}, D.},
        title = "{The Sardinia Radio Telescope Metrology System}",
    booktitle = {2023 XXXVth General Assembly and Scientific Symposium of the International Union of Radio Science (URSI GASS},
         year = 2023,
        month = oct,
          eid = {135},
        pages = {135},
          doi = {10.23919/URSIGASS57860.2023.10265371},
       adsurl = {https://ui.adsabs.harvard.edu/abs/2023ursi.confE.135A}
}

@ARTICLE{muders06,
       author = {{Muders}, D. and {Hafok}, H. and {Wyrowski}, F. and {Polehampton}, E. and {Belloche}, A. and {K{\"o}nig}, C. and {Schaaf}, R. and {Schuller}, F. and {Hatchell}, J. and {van der Tak}, F.},
        title = "{APECS - the Atacama pathfinder experiment control system}",
      journal = {\aap},
         year = 2006,
        month = aug,
       volume = {454},
       number = {2},
        pages = {L25-L28},
          doi = {10.1051/0004-6361:20065359},
archivePrefix = {arXiv},
       eprint = {astro-ph/0605128},
 primaryClass = {astro-ph},
       adsurl = {https://ui.adsabs.harvard.edu/abs/2006A&A...454L..25M}
}

@ARTICLE{2025ORE.....4..113D,
       author = {{Di Mascolo}, Luca and {Perrott}, Yvette and {Mroczkowski}, Tony and {Raghunathan}, Srinivasan and {Andreon}, Stefano and {Ettori}, Stefano and {Simionescu}, Aurora and {van Marrewijk}, Joshiwa and {Cicone}, Claudia and {Lee}, Minju and {Nelson}, Dylan and {Sommovigo}, Laura and {Booth}, Mark and {Klaassen}, Pamela and {Andreani}, Paola and {Cordiner}, Martin A. and {Johnstone}, Doug and {van Kampen}, Eelco and {Liu}, Daizhong and {Maccarone}, Thomas J. and {Morris}, Thomas W. and {Orlowski-Scherer}, John and {Saintonge}, Am{\'e}lie and {Smith}, Matthew and {Thelen}, Alexander E. and {Wedemeyer}, Sven},
        title = "{Atacama Large Aperture Submillimeter Telescope (AtLAST) science: Resolving the hot and ionized Universe through the Sunyaev-Zeldovich effect}",
      journal = {Open Research Europe},
         year = 2025,
        month = jun,
       volume = {4},
          eid = {113},
        pages = {113},
          doi = {10.12688/openreseurope.17449.2},
archivePrefix = {arXiv},
       eprint = {2403.00909},
 primaryClass = {astro-ph.CO},
       adsurl = {https://ui.adsabs.harvard.edu/abs/2025ORE.....4..113D}
}

@ARTICLE{2025ORE.....4..148L,
       author = {{Liu}, Daizhong and {Saintonge}, Amelie and {Bot}, Caroline and {Kemper}, Francisca and {Lopez-Rodriguez}, Enrique and {Smith}, Matthew and {Stanke}, Thomas and {Andreani}, Paola and {Boselli}, Alessandro and {Cicone}, Claudia and {Davis}, Timothy A. and {Hagedorn}, Bendix and {Lasrado}, Akhil and {Mao}, Ann and {Viti}, Serena and {Booth}, Mark and {Klaassen}, Pamela and {Mroczkowski}, Tony and {Bigiel}, Frank and {Chevance}, Melanie and {Cordiner}, Martin A. and {Di Mascolo}, Luca and {Johnstone}, Doug and {Lee}, Minju and {Maccarone}, Thomas and {Thelen}, Alexander E. and {van Kampen}, Eelco and {Wedemeyer}, Sven},
        title = "{Atacama Large Aperture Submillimeter Telescope (AtLAST) science: Gas and dust in nearby galaxies}",
      journal = {Open Research Europe},
         year = 2025,
        month = may,
       volume = {4},
          eid = {148},
        pages = {148},
          doi = {10.12688/openreseurope.17459.2},
archivePrefix = {arXiv},
       eprint = {2403.01202},
 primaryClass = {astro-ph.GA},
       adsurl = {https://ui.adsabs.harvard.edu/abs/2025ORE.....4..148L}
}

@ARTICLE{2025ORE.....4..132O,
       author = {{Orlowski-Scherer}, John and {Maccarone}, Thomas and {Bright}, Joe and {Kami{\'n}ski}, Tomasz and {Koss}, Michael and {Mohan}, Atul and {Miguel Montenegro-Montes}, Francisco and {N{\ae}ss}, Sigurd and {Ricci}, Claudio and {Severgnini}, Paola and {Stanke}, Thomas and {Vignali}, Cristian and {Wedemeyer}, Sven and {Booth}, Mark and {Cicone}, Claudia and {Di Mascolo}, Luca and {Johnstone}, Doug and {Mroczkowski}, Tony and {Cordiner}, Martin and {Greiner}, Jochen and {Hatziminaoglou}, Evanthia and {van Kampen}, Eelco and {Klaassen}, Pamela and {Lee}, Minju and {Liu}, Daizhong and {Saintonge}, Am{\'e}lie and {Smith}, Matthew and {Thelen}, Alexander},
        title = "{Atacama Large Aperture Submillimeter Telescope (AtLAST) science: Probing the transient and time-variable sky}",
      journal = {Open Research Europe},
         year = 2025,
        month = jan,
       volume = {4},
          eid = {132},
        pages = {132},
          doi = {10.12688/openreseurope.17686.2},
archivePrefix = {arXiv},
       eprint = {2404.13133},
 primaryClass = {astro-ph.CO},
       adsurl = {https://ui.adsabs.harvard.edu/abs/2025ORE.....4..132O}
}

@ARTICLE{2024ORE.....4..140W,
       author = {{Wedemeyer}, Sven and {Barta}, Miroslav and {Braj{\v{s}}a}, Roman and {Chai}, Yi and {Costa}, Joaquim and {Gary}, Dale and {Gimenez de Castro}, Guillermo and {Gunar}, Stanislav and {Fleishman}, Gregory and {Hales}, Antonio and {Hudson}, Hugh and {Kirkaune}, Mats and {Mohan}, Atul and {Motorina}, Galina and {Pellizzoni}, Alberto and {Saberi}, Maryam and {Selhorst}, Caius L. and {Simoes}, Paulo J.~A. and {Shimojo}, Masumi and {Skoki{\'c}}, Ivica and {Sudar}, Davor and {Menezes}, Fabian and {White}, Stephen M. and {Booth}, Mark and {Klaassen}, Pamela and {Cicone}, Claudia and {Mroczkowski}, Tony and {Cordiner}, Martin A. and {Di Mascolo}, Luca and {Johnstone}, Doug and {van Kampen}, Eelco and {Lee}, Minju and {Liu}, Daizhong and {Maccarone}, Thomas and {Orlowski-Scherer}, John and {Saintonge}, Amelie and {Smith}, Matthew and {Thelen}, Alexander E.},
        title = "{Science development study for the Atacama Large Aperture Submillimeter Telescope (AtLAST): Solar and stellar observations}",
      journal = {Open Research Europe},
         year = 2024,
        month = jul,
       volume = {4},
          eid = {140},
        pages = {140},
          doi = {10.12688/openreseurope.17453.1},
archivePrefix = {arXiv},
       eprint = {2403.00920},
 primaryClass = {astro-ph.SR},
       adsurl = {https://ui.adsabs.harvard.edu/abs/2024ORE.....4..140W}
}

@ARTICLE{2024ORE.....4..122V,
       author = {{van Kampen}, Eelco and {Bakx}, Tom and {De Breuck}, Carlos and {Chen}, Chian-Chou and {Dannerbauer}, Helmut and {Magnelli}, Benjamin and {Montenegro-Montes}, Francisco Miguel and {Okumura}, Teppei and {Pu}, Sy-Yin and {Rybak}, Matus and {Saintonge}, Amelie and {Cicone}, Claudia and {Hatziminaoglou}, Evanthia and {Hilhorst}, Juli{\"e}tte and {Klaassen}, Pamela and {Lee}, Minju and {Lovell}, Christopher C. and {Lundgren}, Andreas and {Di Mascolo}, Luca and {Mroczkowski}, Tony and {Sommovigo}, Laura and {Booth}, Mark and {Cordiner}, Martin A. and {Ivison}, Rob and {Johnstone}, Doug and {Liu}, Daizhong and {Maccarone}, Thomas J. and {Smith}, Matthew and {Thelen}, Alexander E. and {Wedemeyer}, Sven},
        title = "{Atacama Large Aperture Submillimeter Telescope (AtLAST) science: Surveying the distant Universe}",
      journal = {Open Research Europe},
         year = 2024,
        month = jun,
       volume = {4},
        pages = {122},
          doi = {10.12688/openreseurope.17445.1},
archivePrefix = {arXiv},
       eprint = {2403.02806},
 primaryClass = {astro-ph.CO},
       adsurl = {https://ui.adsabs.harvard.edu/abs/2024ORE.....4..122V}
}

@ARTICLE{2024ORE.....4..117L,
       author = {{Lee}, Minju and {Schimek}, Alice and {Cicone}, Claudia and {Andreani}, Paola and {Popping}, Gergo and {Sommovigo}, Laura and {Appleton}, Philip N. and {Bischetti}, Manuela and {Cantalupo}, Sebastiano and {Chen}, Chian-Chou and {Dannerbauer}, Helmut and {De Breuck}, Carlos and {Di Mascolo}, Luca and {Emonts}, Bjorn H.~C. and {Hatziminaoglou}, Evanthia and {Pensabene}, Antonio and {Rizzo}, Francesca and {Rybak}, Matus and {Shen}, Sijing and {Lundgren}, Andreas and {Booth}, Mark and {Klaassen}, Pamela and {Mroczkowski}, Tony and {Cordiner}, Martin A. and {Johnstone}, Doug and {van Kampen}, Eelco and {Liu}, Daizhong and {Maccarone}, Thomas and {Saintonge}, Amelie and {Smith}, Matthew and {Thelen}, Alexander E. and {Wedemeyer}, Sven},
        title = "{Atacama Large Aperture Submillimeter Telescope (AtLAST) science: The hidden circumgalactic medium}",
      journal = {Open Research Europe},
         year = 2024,
        month = jun,
       volume = {4},
        pages = {117},
          doi = {10.12688/openreseurope.17452.1},
archivePrefix = {arXiv},
       eprint = {2403.00924},
 primaryClass = {astro-ph.GA},
       adsurl = {https://ui.adsabs.harvard.edu/abs/2024ORE.....4..117L}
}

@ARTICLE{2024ORE.....4..112K,
       author = {{Klaassen}, Pamela and {Traficante}, Alessio and {Beltr{\'a}n}, Maria and {Pattle}, Kate and {Booth}, Mark and {Lovell}, Joshua and {Marshall}, Jonathan and {Hacar}, Alvaro and {Gaches}, Brandt and {Bot}, Caroline and {Peretto}, Nicolas and {Stanke}, Thomas and {Arzoumanian}, Doris and {Duarte Cabral}, Ana and {Duch{\^e}ne}, Gaspard and {Eden}, David and {Hales}, Antonio and {Kauffmann}, Jens and {Luppe}, Patricia and {Marino}, Sebastian and {Redaelli}, Elena and {Rigby}, Andrew and {S{\'a}nchez-Monge}, {\'A}lvaro and {Schisano}, Eugenio and {Semenov}, Dmitry and {Spezzano}, Silvia and {Thompson}, Mark and {Wyrowski}, Friedrich and {Cicone}, Claudia and {Mroczkowski}, Tony and {Cordiner}, Martin and {Di Mascolo}, Luca and {Johnstone}, Doug and {van Kampen}, Eelco and {Lee}, Minju and {Liu}, Daizhong and {Maccarone}, Thomas and {Saintonge}, Am{\'e}lie and {Smith}, Matthew and {Thelen}, Alexander and {Wedemeyer}, Sven},
        title = "{Atacama Large Aperture Submillimeter Telescope (AtLAST) science: Our Galaxy}",
      journal = {Open Research Europe},
         year = 2024,
        month = jun,
       volume = {4},
        pages = {112},
          doi = {10.12688/openreseurope.17450.1},
archivePrefix = {arXiv},
       eprint = {2403.00917},
 primaryClass = {astro-ph.GA},
       adsurl = {https://ui.adsabs.harvard.edu/abs/2024ORE.....4..112K}
}

@ARTICLE{2024ORE.....4...78C,
       author = {{Cordiner}, Martin and {Thelen}, Alexander and {Cavalie}, Thibault and {Cosentino}, Richard and {Fletcher}, Leigh N. and {Gurwell}, Mark and {de Kleer}, Katherine and {Kuan}, Yi-Jehng and {Lellouch}, Emmanuel and {Moullet}, Arielle and {Nixon}, Conor and {de Pater}, Imke and {Teanby}, Nicholas and {Butler}, Bryan and {Charnley}, Steven and {Milam}, Stefanie and {Moreno}, Raphael and {Booth}, Mark and {Klaassen}, Pamela and {Cicone}, Claudia and {Mroczkowski}, Tony and {Di Mascolo}, Luca and {Johnstone}, Doug and {van Kampen}, Eelco and {Lee}, Minju and {Liu}, Daizhong and {Maccarone}, Thomas and {Saintonge}, Amelie and {Smith}, Matthew and {Wedemeyer}, Sven},
        title = "{Atacama Large Aperture Submillimeter Telescope (AtLAST) science: Planetary and cometary atmospheres}",
      journal = {Open Research Europe},
         year = 2024,
        month = apr,
       volume = {4},
        pages = {78},
          doi = {10.12688/openreseurope.17473.1},
archivePrefix = {arXiv},
       eprint = {2403.02258},
 primaryClass = {astro-ph.EP},
       adsurl = {https://ui.adsabs.harvard.edu/abs/2024ORE.....4...78C}
}

@ARTICLE{Groppi2019,
       author = {{Groppi}, Christopher and {Baryshev}, Andrey and {Graf}, Urs and {Wiedner}, Martina and {Klaassen}, Pamela and {Mroczkowski}, Tony},
        title = "{First Generation Heterodyne Instrumentation Concepts for the Atacama Large Aperture Submillimeter Telescope}",
      journal = {arXiv e-prints},
         year = 2019,
        month = jul,
          eid = {arXiv:1907.03479},
        pages = {arXiv:1907.03479},
archivePrefix = {arXiv},
       eprint = {1907.03479},
 primaryClass = {astro-ph.IM},
       adsurl = {https://ui.adsabs.harvard.edu/abs/2019arXiv190703479G},
 howpublished = "\url{https://www.nrao.edu/meetings/isstt/2019.shtml}",
         note = "\url{https://www.nrao.edu/meetings/isstt/2019.shtml}"
}

@ARTICLE{AtLAST_memo_3,
       author = {{van Kampen}, E.},
        title = "{Instrumentation WG memo on instrument mounting options}",
      journal = {AtLAST Memo Series},
       volume = 3,
         year = 2022,
        month = Dec,
 howpublished = "\url{https://atlast-telescope.org/publications/memo-series/memo-public/memo3_instrument_mounting_options.pdf}",
         note = "\url{https://atlast-telescope.org/publications/memo-series/memo-public/memo3_instrument_mounting_options.pdf}"
}

@ARTICLE{AtLAST_memo_9,
       author = {{De Breuck}, C. and {Otarola}, A. {Pérez Beaupuits},  J.P. and
{Nyman}, L-\AA and {Valenzuela}, G. and {Reichert}, M.},
        title = "{AtLAST site selection report}",
      journal = {AtLAST Memo Series},
       volume = 9,
         year = 2024,
        month = Aug,
 howpublished = "\url{https://atlast-telescope.org/publications/memo-series/memo-public/d3.2_site_selection_report_final.pdf}",
         note = "\url{https://atlast-telescope.org/publications/memo-series/memo-public/d3.2_site_selection_report_final.pdf}"
}

@ARTICLE{Kovacs2025,
       author = {{Kov{\'a}cs}, Attila and {Keating}, Garrett K. and {Greve}, Thomas R. and {Norton}, Timothy},
        title = "{Concept integral field unit spectrometer instrument for the next-generation millimeter-wave cosmological surveys}",
      journal = {Journal of Astronomical Telescopes, Instruments, and Systems},
         year = 2025,
        month = oct,
       volume = {11},
          eid = {045007},
        pages = {045007},
          doi = {10.1117/1.JATIS.11.4.045007},
       adsurl = {https://ui.adsabs.harvard.edu/abs/2025JATIS..11d5007K}
}

@ARTICLE{Zhu2021,
       author = {{Zhu}, Ningfeng and {Bhandarkar}, Tanay and {Coppi}, Gabriele and {Kofman}, Anna M. and {Orlowski-Scherer}, John L. and et al.},
        title = "{The Simons Observatory Large Aperture Telescope Receiver}",
      journal = {\apjs},
         year = 2021,
        month = sep,
       volume = {256},
       number = {1},
          eid = {23},
        pages = {23},
          doi = {10.3847/1538-4365/ac0db7},
archivePrefix = {arXiv},
       eprint = {2103.02747},
 primaryClass = {astro-ph.IM},
       adsurl = {https://ui.adsabs.harvard.edu/abs/2021ApJS..256...23Z}
}

@ARTICLE{Akiyama2023,
       author = {{Akiyama}, Kazunori and {Kauffmann}, Jens and {Matthews}, Lynn D. and {Moriyama}, Kotaro and {Koyama}, Shoko and {Hada}, Kazuhiro},
        title = "{Millimeter/Submillimeter VLBI with a Next Generation Large Radio Telescope in the Atacama Desert}",
      journal = {Galaxies},
         year = 2023,
        month = jan,
       volume = {11},
       number = {1},
          eid = {1},
        pages = {1},
          doi = {10.3390/galaxies11010001},
archivePrefix = {arXiv},
       eprint = {2212.05118},
 primaryClass = {astro-ph.IM},
       adsurl = {https://ui.adsabs.harvard.edu/abs/2023Galax..11....1A}
}

@ARTICLE{Moerman2026,
       author = {{Moerman}, A. and {Soshnin}, N. and {Brackenhoff}, S.~A. and {Dabironezare}, S.~O. and {Karatsu}, K. and {Marting}, L.~H. and {de Rooij}, S.~A.~H. and {Roos}, M. and {Brandl}, B.~R. and {Endo}, A.},
        title = "{gateau: an observation simulator for ground-based submillimeter astronomy with integral field units and kinetic inductance detectors}",
      journal = {arXiv e-prints},
         year = 2026,
        month = apr,
          eid = {arXiv:2604.23305},
        pages = {arXiv:2604.23305},
          doi = {10.48550/arXiv.2604.23305},
archivePrefix = {arXiv},
       eprint = {2604.23305},
       volume = {2604.23305},
 primaryClass = {astro-ph.IM},
       adsurl = {https://ui.adsabs.harvard.edu/abs/2026arXiv260423305M}
}

@book{Dryzek2022,
author = {Dryzek, John S.},
address = {Oxford},
booktitle = {The politics of the Earth : environmental discourses},
edition = {Fourth edition.},
isbn = {9780198851745},
language = {eng},
publisher = {Oxford University Press},
title = {The politics of the Earth : environmental discourses },
year = {2022},
}

@BOOK{Wilson2013,
       author = {{Wilson}, Thomas L. and {Rohlfs}, Kristen and {H{\"u}ttemeister}, Susanne},
        title = "{Tools of Radio Astronomy}",
         year = 2013,
          doi = {10.1007/978-3-642-39950-3},
       adsurl = {https://ui.adsabs.harvard.edu/abs/2013tra..book.....W}
}

@ARTICLE{Dunner2013,
       author = {{D{\"u}nner}, Rolando and {Hasselfield}, Matthew and {Marriage}, Tobias A. and {Sievers}, Jon and {Acquaviva}, Viviana and {Addison}, Graeme E. and et al.},
        title = "{The Atacama Cosmology Telescope: Data Characterization and Mapmaking}",
      journal = {\apj},
         year = 2013,
        month = jan,
       volume = {762},
       number = {1},
          eid = {10},
        pages = {10},
          doi = {10.1088/0004-637X/762/1/10},
archivePrefix = {arXiv},
       eprint = {1208.0050},
 primaryClass = {astro-ph.IM},
       adsurl = {https://ui.adsabs.harvard.edu/abs/2013ApJ...762...10D}
}
\bibliographystyle{spiebib} 

\end{document}